\documentclass[twocolumn,aip,jcp,10pt,superscriptaddress,longbibliography,floatfix,citeautoscript]{revtex4-2}

\usepackage{amsmath}
\usepackage{amssymb}
\usepackage{bm}
\usepackage{dcolumn}
\usepackage{glossaries}
\usepackage{graphicx}
\usepackage{multirow}
\usepackage{fancyvrb}
\usepackage[linesnumbered,ruled]{algorithm2e}
\usepackage{siunitx}

\usepackage[usenames,dvipsnames,svgnames]{xcolor}
\usepackage{hyperref}
\hypersetup{
    pdfnewwindow=true,      
    colorlinks=true,        
    linkcolor=Blue,         
    citecolor=Blue,         
    filecolor=Blue,         
    urlcolor=Blue           
}

\usepackage{listings}	
\newacronym{ace}{ACE}{atomic cluster expansion}
\newacronym{al}{AL}{active learning}
\newacronym{dft}{DFT}{density functional theory}
\newacronym{dp}{DP}{deep potential}
\newacronym{eam}{EAM}{embedded atom method}
\newacronym{gap}{GAP}{Gaussian approximation potential}
\newacronym{hnemd}{HNEMD}{homogeneous non-equilibrium molecular dynamics}
\newacronym{lda}{LDA}{local density approximation}
\newacronym{mae}{MAE}{mean absolute error}
\newacronym{md}{MD}{molecular dynamics}
\newacronym{mlp}{MLP}{machine-learned potential}
\newacronym{mtp}{MTP}{moment tensor potential}
\newacronym{nep}{NEP}{neuroevolution potential}
\newacronym{pc}{PC}{principal component}
\newacronym{reann}{REANN}{recursive embedded-atom neural network}
\newacronym{rmse}{RMSE}{root mean square error}
\newacronym{snes}{SNES}{separable natural evolution strategy}
\newacronym{soap}{SOAP}{smooth overlap of atomic positions}
\newacronym{vacf}{VACF}{velocity auto-correlation function}
\newacronym{vdos}{VDOS}{vibrational density of states}
\newacronym{vdw}{vdW}{van-der-Waals}

\DeclareSIUnit\angstrom{\text{Å}}
\DeclareSIUnit{\atom}{atom}
\DeclareSIUnit{\step}{step}
\DeclareSIUnit{\atomstepsecond}{\atom\step\per\second}

\newcolumntype{d}{D{.}{.}{-1}}

\begin{document}

\title{GPUMD: A package for constructing accurate machine-learned potentials and performing highly efficient atomistic simulations}


\author{Zheyong Fan}
\email{brucenju@gmail.com}
\affiliation{College of Physical Science and Technology, Bohai University, Jinzhou 121013, P. R. China}

\author{Yanzhou Wang}
\affiliation{MSP group, QTF Centre of Excellence, Department of Applied Physics, Aalto University, FI-00076 Aalto, Espoo, Finland}
\affiliation{Beijing Advanced Innovation Center for Materials Genome Engineering,  University of Science and Technology Beijing, Beijing 100083, China}

\author{Penghua Ying}
\affiliation{School of Science, Harbin Institute of Technology, Shenzhen, 518055, P. R. China}

\author{Keke Song}
\affiliation{Beijing Advanced Innovation Center for Materials Genome Engineering,  University of Science and Technology Beijing, Beijing 100083, China}

\author{Junjie Wang}
\affiliation{National Laboratory of Solid State Microstructures, School of Physics and Collaborative Innovation Center of Advanced Microstructures, Nanjing University, Nanjing 210093, China}

\author{Yong Wang}
\affiliation{National Laboratory of Solid State Microstructures, School of Physics and Collaborative Innovation Center of Advanced Microstructures, Nanjing University, Nanjing 210093, China}

\author{Zezhu Zeng}
\affiliation{Department of Mechanical Engineering, The University of Hong Kong, Pokfulam Road, Hong Kong SAR, China}

\author{Ke Xu}
\affiliation{Department of Physics, Research Institute for Biomimetics and Soft Matter, Jiujiang Research Institute and Fujian Provincial Key Laboratory for Soft Functional Materials Research, Xiamen University, Xiamen 361005, PR China.}

\author{Eric Lindgren}
\affiliation{
  Chalmers University of Technology,
  Department of Physics,
  41926 Gothenburg, Sweden
}

\author{J. Magnus Rahm}
\affiliation{
  Chalmers University of Technology,
  Department of Physics,
  41926 Gothenburg, Sweden
}

\author{Alexander J. Gabourie}
\affiliation{Department of Electrical Engineering, Stanford University, Stanford, California 94305, USA}

\author{Jiahui Liu}
\affiliation{Beijing Advanced Innovation Center for Materials Genome Engineering,  University of Science and Technology Beijing, Beijing 100083, China}

\author{Haikuan Dong}
\affiliation{College of Physical Science and Technology, Bohai University, Jinzhou 121013, P. R. China}
\affiliation{Beijing Advanced Innovation Center for Materials Genome Engineering,  University of Science and Technology Beijing, Beijing 100083, China}

\author{Jianyang Wu}
\affiliation{Department of Physics, Research Institute for Biomimetics and Soft Matter, Jiujiang Research Institute and Fujian Provincial Key Laboratory for Soft Functional Materials Research, Xiamen University, Xiamen 361005, PR China.}

\author{Yue Chen}
\affiliation{Department of Mechanical Engineering, The University of Hong Kong, Pokfulam Road, Hong Kong SAR, China}

\author{Zheng Zhong}
\affiliation{School of Science, Harbin Institute of Technology, Shenzhen, 518055, P. R. China}

\author{Jian Sun}
\email{jiansun@nju.edu.cn}
\affiliation{National Laboratory of Solid State Microstructures, School of Physics and Collaborative Innovation Center of Advanced Microstructures, Nanjing University, Nanjing 210093, China}

\author{Paul Erhart}
\email{erhart@chalmers.se}
\affiliation{
  Chalmers University of Technology,
  Department of Physics,
  41926 Gothenburg, Sweden
}

\author{Yanjing Su}
\email{yjsu@ustb.edu.cn}
\affiliation{Beijing Advanced Innovation Center for Materials Genome Engineering,  University of Science and Technology Beijing, Beijing 100083, China}

\author{Tapio Ala-Nissila}
\affiliation{MSP group, QTF Centre of Excellence, Department of Applied Physics, Aalto University, FI-00076 Aalto, Espoo, Finland}
\affiliation{Interdisciplinary Centre for Mathematical Modelling, Department of Mathematical Sciences, Loughborough University, Loughborough, Leicestershire LE11 3TU, UK}

\date{\today}

\begin{abstract}
We present our latest advancements of machine-learned potentials (MLPs) based on the neuroevolution potential (NEP) framework introduced in [Fan \textit{et al.}, Phys. Rev. B \textbf{104}, 104309 (2021)] and their implementation in the open-source package \textsc{gpumd}.
We increase the accuracy of NEP models both by improving the radial functions in the atomic-environment descriptor using a linear combination of Chebyshev basis functions and by extending the angular descriptor with some four-body and five-body contributions as in the atomic cluster expansion approach.
We also detail our efficient implementation of the NEP approach in graphics processing units as well as our workflow for the construction of NEP models, and we demonstrate their application in large-scale atomistic simulations.
By comparing to state-of-the-art MLPs, we show that the NEP approach not only achieves above-average accuracy but also is far more computationally efficient.
These results demonstrate that the \textsc{gpumd} package is a promising tool for solving challenging problems requiring highly accurate, large-scale atomistic simulations.
To enable the construction of MLPs using a minimal training set, we propose an active-learning scheme based on the latent space of a pre-trained NEP model.
Finally, we introduce three separate Python packages, \textsc{gpyumd}, \textsc{calorine}, and \textsc{pynep}, which enable the integration of \textsc{gpumd} into Python workflows.
\end{abstract}

\maketitle

\section{Introduction}

Machine-learned classical potentials \cite{behler2016jcp,Deringer2019am,Mueller2020jcp,Noe2020arpc,Miksch2021mlst,Mishin2021am,Unke2021cr} have shown great promise in enabling accurate atomistic simulations far beyond the space and time scales that can be achieved using quantum mechanical calculations. Many open-source computer packages for \glspl{mlp} have been published, including \textsc{quip-gap} \cite{bartok2010prl, bartok2013prb, Deringer2017prb}, \textsc{snap} \cite{Thompson2014jcp}, \textsc{amp} \cite{Khorshidi2016cpc}, \textsc{aenet}\cite{Artrith2016cms, Artrith2017prb}, \textsc{ani} \cite{smith2017cs}, \textsc{SchNet} \cite{Schutt2018jcp}, \textsc{DeepMD-kit} \cite{wang2018cpc,zhang2018prl,zhang2018endtoend}, \textsc{TensorMol}\cite{yao2018cs}, \textsc{PhysNet} \cite{Unke2019jctc}, \textsc{MEGNet} \cite{chen2019cm}, \textsc{turboGAP} \cite{Caro2019prb}, \textsc{sGDML} \cite{Chmiela2019cpc}, \textsc{n2p2} \cite{Singraber2019jctc}, \textsc{simple-nn} \cite{lee2019cpc}, \textsc{panna} \cite{lot2020cpc}, \textsc{fchl} \cite{Christensen2020jcp}, \textsc{PiNN} \cite{shao2020jcim}, \textsc{mlip} \cite{Shapeev2016,Novikov2021}, \textsc{reann} \cite{zhang2022jcp}, \textsc{tabGAP} \cite{Byggmastar2021prb}, \textsc{pyXtal-FF} \cite{Yanxon2021}, \textsc{python-ace} \cite{lysogorskiy2021npj,bochkarev2022prm}, and \textsc{kliff} \cite{wen2022cpc}.
However, most existing implementations of \glspl{mlp} (\textsc{tabGAP} \cite{Byggmastar2021prb} is a notable exception) have a computational speed of about \SI{1e3}{\atomstepsecond} using one typical CPU core, which is about two to three orders of magnitude slower than typical empirical potentials such as the Tersoff potential \cite{tersoff1989prb}.
Therefore, even though \glspl{mlp} are already faster than quantum-mechanical methods, it is still desirable to speed up \glspl{mlp} as much as possible. 

One approach to speed up \glspl{mlp} is to use a huge number of CPUs and/or GPUs through message-passing information parallelisation.
For example, using 27,360 V100 GPUs and the same number of CPU cores, the \gls{dp} approach \cite{wang2018cpc, zhang2018prl, zhang2018endtoend} has been used to simulate a 127-million-atom aluminum system with a speed of about \SI{1.23e9}{\atomstepsecond} \cite{jia2020GB}.
However, such huge amounts of computational resources are not available to most researchers.
More importantly, the performance of \gls{dp} per V100 GPU is only about \SI{4.5e4}{\atomstepsecond}, which is about ten times \textit{slower} than an empirical \gls{eam} potential \cite{daw1984prb} with a single typical CPU core.

A more economical approach is to optimize the formulation and implementation of the \gls{mlp} itself so that it can attain a high computational speed using a reasonable amount of computational resources available to most researchers.
To this end, we have developed a \gls{mlp} called \gls{nep} \cite{fan2021neuroevolution, fan2022jpcm} within the \textsc{gpumd} package \cite{fan2013cpc, fan2017cpc} that can achieve a computational speed of about \SI{1e7}{\atomstepsecond} using a single V100 GPU, which is about ten times \textit{faster} than the empirical Tersoff potential on a single typical CPU core. 

In this paper, we present recent developments of the NEP approach that further improve its accuracy without reducing the efficiency.
Specifically, we improve the radial functions in the atomic-environment descriptor by using a combination of basis functions, and add angular descriptor components with high-order correlations.
The improved radial functions are better at distinguishing different atom types and lead to higher accuracy in multi-components systems.
The added angular descriptor components with high-order correlations make the atomic-environment descriptor more complete and help to increase the regression accuracy.

Using a number of systems, including MgAlCu alloy, silicon with various phases, the azobenzene molecule, and carbon with various phases, we demonstrate the accuracy and efficiency of the latest NEP as implemented in the \textsc{gpumd} package.
We compare with other state-of-the-art \glspl{mlp}, including \gls{dp} \cite{wang2018cpc,zhang2018prl,zhang2018endtoend}, \gls{gap} \cite{bartok2010prl}, \gls{mtp} \cite{Shapeev2016,Novikov2021}, \gls{reann} \cite{zhang2021prl,zhang2022jcp}, and \gls{ace} \cite{drautz2019prb,kovacs2021jctc,Dusson2021jcp}.
Through these comprehensive comparisons, we show that the NEP implementation in \textsc{gpumd} can achieve a computational speed that is far superior to other \glspl{mlp}, under the condition of achieving an above-average accuracy.
We present the algorithms for the efficient GPU implementation of NEP in great detail.
Using a single GPU, such as an A100, one can use \textsc{gpumd} to simulate up to 10 million atoms on nanosecond time scales, which can only be achieved by using a huge amount of computational resources with other publicly available codes.
The \textsc{gpumd} package makes large-scale, high-accuracy atomistic simulations available to a wide community instead of only a small number of institutions.
In addition to these high-efficiency atomistic simulations, we also propose an effective active-learning scheme based on the latent space of the NEP model that can greatly reduce the computational burden of preparing training data.

Finally, we introduce the workflow for constructing and using NEPs through concrete examples in atomistic simulations of various materials properties, including lattice constant, elastic constants, stress-strain relation during tensile loading, structural properties during a melt-quench-anneal process, and thermal properties of amorphous structures.
We also describe interfacing \textsc{gpumd} to Python via the \textsc{gpyumd}, \textsc{calorine}, and \textsc{pynep} packages.

\section{Theoretical formulations of the NEP approach}

The first NEP, called NEP1, was proposed in Ref.~\onlinecite{fan2021neuroevolution}.
An improved version, called NEP2, was presented in Ref.~\onlinecite{fan2022jpcm}.
In the present paper, we further refine the NEP approach and introduce NEP3.
In this section, we present NEP3 and discuss the differences to NEP1 and NEP2.

\subsection{The neural-network model}

Following Behler and Parrinello \cite{behler2007prl}, the site energy of atom $i$ is taken as a function of the  descriptor vector with $N_\mathrm{des}$ components,
$
U_i(\mathbf{q}) = U_i \left(\{q^i_{\nu}\}_{\nu =1}^{N_\mathrm{des}}\right)
$.
We use a feedforward neural-network with a single hidden layer with $N_\mathrm{neu}$ neurons to model this function:
\begin{equation}
\label{equation:Ui}
U_i = \sum_{\mu=1}^{N_\mathrm{neu}}w^{(1)}_{\mu}\tanh\left(\sum_{\nu=1}^{N_\mathrm{des}} w^{(0)}_{\mu\nu} q^i_{\nu} - b^{(0)}_{\mu}\right) - b^{(1)},
\end{equation}
where $\tanh(x)$ is the activation function in the hidden layer, $\mathbf{w}^{(0)}$ is the connection weight matrix from the input layer (descriptor vector) to the hidden layer, $\mathbf{w}^{(1)}$ is the connection weight vector from the hidden layer to the output node, which is the energy $U_i$, $\mathbf{b}^{(0)}$ is the bias vector in the hidden layer, and $b^{(1)}$ is the bias for node $U_i$.
The total number of parameters in the neural network is thus $(N_\mathrm{des}+2)N_\mathrm{neu}+1$.
The descriptor vector is formed by juxtaposition of a number of components, including those with radial (distance) information only, which are called radial descriptor components, and those with both radial and angular information, which are called angular descriptor components.
The descriptor is one of the most important aspects in \glspl{mlp} \cite{Musil2021cr,Langer2022npj}. We discuss the radial and angular descriptor components below.

\subsection{Radial descriptor components}

There are $n_\mathrm{max}^\mathrm{R}+1$ radial descriptor components and they are defined as
\begin{equation}
\label{equation:qin}
q^i_{n}
= \sum_{j\neq i} g_{n}(r_{ij})
\quad\text{with}\quad
0\leq n\leq n_\mathrm{max}^\mathrm{R},
\end{equation}
where the summation runs over all the neighbors of atom $i$ within a certain cutoff distance.
The functions $g_n(r_{ij})$ depend on the distance $r_{ij}$ only and are therefore called the radial functions.
In NEP3, they are defined as a linear combination of $N_\mathrm{bas}^\mathrm{R}+1$ basis functions $\{f_k(r_{ij})\}_{k=0}^{N_\mathrm{bas}^\mathrm{R}}$:
\begin{align}
g_n(r_{ij}) &= \sum_{k=0}^{N_\mathrm{bas}^\mathrm{R}} c^{ij}_{nk} f_k(r_{ij}),\quad\text{with}
\label{equation:g_n}
\\
f_k(r_{ij}) &= \frac{1}{2}
\left[
    T_k\left(2\left(r_{ij}/r_\mathrm{c}^\mathrm{R}-1\right)^2-1\right)+1
\right]
f_\mathrm{c}(r_{ij}).
\label{equation:f_n}
\end{align}
Both $n_\mathrm{max}^\mathrm{R}$ and $N_\mathrm{bas}^\mathrm{R}$ are tunable hyperparameters in NEP3 which, along with other ones, will be listed in \autoref{section:train_nep}.
In \autoref{section:compare}, we will use a few examples to illustrate the judicious choice of the various hyperparameters in typical applications.
Here, $T_k(x)$ is the $k^{\rm th}$ order Chebyshev polynomial of the first kind and $f_\mathrm{c}(r_{ij})$ is the cutoff function defined as
\begin{equation}
   f_\mathrm{c}(r_{ij}) 
   = \begin{cases}
   \frac{1}{2}\left[
   1 + \cos\left( \pi \frac{r_{ij}}{r_\mathrm{c}^\mathrm{R}} \right) 
   \right],& r_{ij}\leq r_\mathrm{c}^\mathrm{R}; \\
   0, & r_{ij} > r_\mathrm{c}^\mathrm{R}.
   \end{cases}
\end{equation}
Here, $r_\mathrm{c}^\mathrm{R}$ is the cutoff distance of the radial descriptor components.
The expansion coefficients $c_{nk}^{ij}$ depend on $n$ and $k$ and also on the types of atoms $i$ and $j$.
Due to the summation over neighbors, the radial descriptor components defined above are invariant with respect to permutation of atoms of the same type. 

If the material considered has $N_\mathrm{typ}$ atom types, the number of $c_{nk}^{ij}$ coefficients for the radial descriptor components is $N_\mathrm{typ}^2 \left(n_\mathrm{max}^\mathrm{R} + 1\right)\left(N_\mathrm{bas}^\mathrm{R} + 1\right)$.
In NEP2, each radial function is simply a basis function and $c_{nk}^{ij}$ is reduced to $\delta_{nk}c_{nij}$, where $\delta_{nk}$ is the Kronecker symbol and $c_{nij}$ is defined in Ref. \onlinecite{fan2022jpcm}.
In NEP2 and NEP3, these coefficients are trainable (similar to the parameters in the neural network), while in NEP1, we have used fixed coefficients similar to previous works \cite{Artrith2017prb, Gastegger2018jcp}.
In Ref.~\onlinecite{fan2022jpcm}, we have shown that NEP2 is much more accurate than NEP1 for multi-component systems.
In this paper, we will show that the accuracy of NEP3 for multi-component systems is further improved as compared to NEP2. 

\subsection{Angular descriptor components}

In NEP1 and NEP2 the angular descriptor components $\{q^i_{nl}\}$ are taken as ($0\leq n\leq n_\mathrm{max}^\mathrm{A}$ and $1\leq l \leq l_\mathrm{max}^\mathrm{3b}$):
\begin{equation}
q^i_{nl} 
= \frac{2l+1}{4\pi}\sum_{j\neq i}\sum_{k\neq i} g_n(r_{ij}) g_n(r_{ik})
P_l(\cos\theta_{ijk}),
\label{equation:qinl_legendre}
\end{equation}
where $P_l(\cos\theta_{ijk})$ is the Legendre polynomial of order $l$ and $\theta_{ijk}$ is the angle formed by the $ij$ and $ik$ bonds.
The radial functions $g_n(r_{ij})$ have the same forms as in Eq.~\eqref{equation:g_n}, but can have a different cutoff distance $r_\mathrm{c}^\mathrm{A}$ and a different basis size $N_\mathrm{bas}^\mathrm{A}$ than those in the radial descriptor components.
Usually, it is beneficial to use $r_\mathrm{c}^\mathrm{A} < r_\mathrm{c}^\mathrm{R}$, assuming that there is no directional dependence of the descriptor on some neighboring atoms that are sufficiently far away from the central atom.
This reflects the physical intuition that interaction strength decreases both with distance and order.
The radial descriptor components are relatively cheap to evaluate and one can thus use a relatively large radial cutoff distance $r_\mathrm{c}^\mathrm{R}$ (also relatively large $n_\mathrm{max}^\mathrm{R}$ and $N_\mathrm{bas}^\mathrm{R}$) combined with a relatively small angular cutoff distance $r_\mathrm{c}^\mathrm{A}$ (also relatively small $n_\mathrm{max}^\mathrm{A}$ and $N_\mathrm{bas}^\mathrm{A}$) to achieve a good balance between accuracy and speed.
Note that message-passing could effectively increase the interaction range but is not necessarily an efficient way of describing long-range interactions \cite{Nigam2022jcp}.
Using a relatively large cutoff for the radial descriptor components is generally a more efficient way of incorporating long-ranged interactions, such as \gls{vdw} and screened Coulomb interactions, although it is incapable of describing genuine long-ranged interactions such as unscreened Coulomb interactions.

The expression \eqref{equation:qinl_legendre} is not efficient for numerical evaluation due to the double summation over neighbors.
An equivalent form that is more efficient for numerical evaluation can be obtained by using the addition theorem of the spherical harmonics as:
\begin{equation}
q^i_{nl} 
= \sum_{m=-l}^l (-1)^m A^i_{nlm} A^i_{nl(-m)} = \sum_{m=0}^l  (2-\delta_{0l}) |A^i_{nlm}|^2,
\label{equation:qinl_spherical}
\end{equation}
where 
\begin{equation}
A^i_{nlm} 
= \sum_{j\neq i} g_n(r_{ij}) Y_{lm}(\theta_{ij},\phi_{ij}).
\label{equation:Ainlm}
\end{equation}
Here, $Y_{lm}(\theta_{ij},\phi_{ij})$ are the spherical harmonics as a function of the polar angle $\theta_{ij}$ and the azimuthal angle $\phi_{ij}$ for the position difference vector $\bm{r}_{ij} \equiv \bm{r}_{j}-\bm{r}_{i}$ from atom $i$ to atom $j$. In Eq.~\eqref{equation:qinl_spherical}, we have used the property $A_{nl(-m)}=(-1)^mA_{nlm}^{\ast}$, which follows from the property $Y_{l(-m)}=(-1)^mY_{lm}^{\ast}$.

The angular descriptor components above are usually known as 3-body ones as in the \gls{ace} approach \cite{drautz2019prb}, although all the descriptor components are many-body in nature.
For simplicity, we will use the \gls{ace} terminology.
Higher-order angular descriptor components can be similarly constructed \cite{drautz2019prb}.
In NEP3, we add the following 4-body descriptor components ($1 \leq l_1 \leq l_2 \leq l_3 \leq l_\mathrm{max}^\mathrm{4b}$):
\begin{align}
q^i_{nl_1l_2l_3} 
=&
\sum_{m_1=-l_1}^{l_1}
\sum_{m_2=-l_2}^{l_2}
\sum_{m_3=-l_3}^{l_3}
\left(
\begin{array}{ccc}
l_1 & l_2 & l_3 \\
m_1 & m_2 & m_3\\
\end{array} 
\right) \nonumber \\
&\times A^{i}_{nl_1m_1}A^{i}_{nl_2m_2}A^{i}_{nl_3m_3},
\label{equation:q_4body}
\end{align}
and the following 5-body ones ($1 \leq l_1 \leq l_2 \leq l_3 \leq l_4 \leq l_\mathrm{max}^\mathrm{5b}$):
\begin{align}
&q^i_{nl_1l_2l_3l_4} 
= 
\sum_{m_1=-l_1}^{l_1}
\sum_{m_2=-l_2}^{l_2}
\sum_{m_3=-l_3}^{l_3}
\sum_{m_4=-l_4}^{l_4}
 \nonumber \\
&\left[
\begin{array}{cccc}
l_1 & l_2 & l_3 & l_4\\
m_1 & m_2 & m_3 & m_4\\
\end{array} 
\right]
 A^{i}_{nl_1m_1}A^{i}_{nl_2m_2}A^{i}_{nl_3m_3}A^{i}_{nl_4m_4},
\label{equation:q_5body}
\end{align}
where 
$
\left(
\begin{array}{ccc}
l_1 & l_2 & l_3 \\
m_1 & m_2 & m_3\\
\end{array} 
\right)
$
are Wigner $3j$ symbols and \cite{drautz2019prb}
\begin{align}
&\left[
\begin{array}{cccc}
l_1 & l_2 & l_3 & l_4\\
m_1 & m_2 & m_3 & m_4\\
\end{array} 
\right]
= \sum_{L=\max\{|l_1-l_2|,|l_3-l_4|\}}^{\min\{|l_1+l_2|,|l_3+l_4|\}}\sum_{M=-L}^L \nonumber \\
&
(-1)^M \left(
\begin{array}{ccc}
L & l_1 & l_2 \\
-M & m_1 & m_2\\
\end{array} 
\right)
\left(
\begin{array}{ccc}
L & l_3 & l_4 \\
M & m_3 & m_4\\
\end{array} 
\right).
\end{align}
We can consider higher-order terms \cite{drautz2019prb}, but to keep a balance between accuracy and speed, we only consider those up to fifth order.
Recent work suggests that 4-body interactions are difficult to learn using only 3-body correlations \cite{Parsaeifard2022jcp} and that using only up to 4-body correlations can still yield identical results for different configurations of simple molecules \cite{Pozdnyakov2021prl}.
One can formally achieve completeness in the \gls{mtp} formalism \cite{Shapeev2016}, the \gls{ace} formalism \cite{drautz2019prb, Dusson2021jcp}, and other related ones \cite{Glielmo2018prb, Willatt2019jcp, Uhrin2021prb}, but in practice, there is always a truncation of the descriptor size and a balance must be struck between accuracy and speed.
This balance is one of the most important guidelines for the development of NEP in \textsc{gpumd}.

\subsection{Explicit expressions for the angular descriptor components}
\label{section:explicit_expressions}

The angular descriptor components are quite complicated and care must be taken to achieve an efficient implementation.
There have been some implementations of the \gls{ace} approach combined with linear regression, where it has been found that recursive evaluation can lead to much higher efficiency \cite{Dusson2021jcp, lysogorskiy2021npj, bochkarev2022prm}.
In our GPU implementation of NEP3 with the \gls{ace}-like descriptor components, we find it crucial to derive the relevant expressions as explicitly as possible as it allows us to reduce the number of terms to be evaluated thanks to symmetry considerations.

To facilitate the following presentation, we define a series of summations that are used to express the angular descriptor components ($0\leq n\leq n_\mathrm{max}^\mathrm{A}$ and $0\leq k\leq 23$): 
\begin{equation}
\label{equation:S_nk}
S_{n,k} = \sum_{j\neq i} \frac{g_n(r_{ij})}{r_{ij}^n} b_k(x_{ij},y_{ij},z_{ij}).
\end{equation}
The functions $b_k(x_{ij},y_{ij},z_{ij})$ here are 
$z_{ij}$, 
$x_{ij}$, 
$y_{ij}$, 
$3z_{ij}^2-r_{ij}^2$, 
$x_{ij} z_{ij}$,
$y_{ij} z_{ij}$,
$x_{ij}^2 - y_{ij}^2$,
$2 x_{ij} y_{ij}$,
$(5 z_{ij}^2 - 3r_{ij}^2) z_{ij}$,
$(5 z_{ij}^2 - r_{ij}^2) x_{ij}$,
$(5 z_{ij}^2 - r_{ij}^2) y_{ij}$,
$(x_{ij}^2 - y_{ij}^2) z_{ij}$,
$2 x_{ij}  y_{ij}  z_{ij}$,
$(x_{ij}^2  - 3 y_{ij}^2) x_{ij}$,
$(3x_{ij}^2  - y_{ij}^2) y_{ij}$,
$(35z_{ij}^2  - 30r_{ij}^2) z_{ij}^2 + 3r_{ij}^4$,
$(7z_{ij}^2  - 3r_{ij}^2) x_{ij}z_{ij}$,
$(7z_{ij}^2  - 3r_{ij}^2) y_{ij}z_{ij}$,
$(7z_{ij}^2  - r_{ij}^2) (x_{ij}^2-y_{ij}^2)$,
$(7z_{ij}^2  - r_{ij}^2) 2x_{ij}y_{ij}$,
$(x_{ij}^2  - 3y_{ij}^2) x_{ij}z_{ij}$,
$(3x_{ij}^2  - y_{ij}^2) y_{ij}z_{ij}$,
$(x_{ij}^2  - y_{ij}^2)^2 - 4x_{ij}^2y_{ij}^2$, and
$4(x_{ij}^2  - y_{ij}^2) x_{ij}y_{ij}$ from $k=0$ to $k=23$.
With these summations, we can write the 3-body angular descriptor components up to $l_\mathrm{max}^\mathrm{3b}=4$ explicitly as:  
\begin{align}
q^i_{n1} = \frac{1}{4}\frac{3}{\pi} S_{n,0}^2  +  2\frac{1}{4}\frac{3}{2\pi} (S_{n,1}^2 + S_{n,2}^2) \equiv \sum_{k=0}^2C^\mathrm{3b}_k S_{n,k}^2;
\end{align}
\begin{align}
q^i_{n2} 
&=\frac{1}{16}\frac{5}{\pi} 
S_{n,3}^2 + 2\frac{1}{4}\frac{15}{2\pi} 
(S_{n,4}^2 + S_{n,5}^2) + 2\frac{1}{16}\frac{15}{2\pi} 
(S_{n,6}^2 + S_{n,7}^2)\nonumber \\
&\equiv \sum_{k=3}^7C^\mathrm{3b}_k S_{n,k}^2;
\end{align}
\begin{align}
q^i_{n3}  
&= \frac{1}{16}\frac{7}{\pi} S_{n,8}^2 
+ 2\frac{1}{64}\frac{21}{\pi} 
(S_{n,9}^2 + S_{n,10}^2) \nonumber \\
&+ 2\frac{1}{16}\frac{105}{2\pi} 
(S_{n,11}^2 + S_{n,12}^2)
+ 2\frac{1}{64}\frac{35}{\pi} 
(S_{n,13}^2 + S_{n,14}^2)\nonumber \\
&\equiv \sum_{k=8}^{14}C^\mathrm{3b}_k S_{n,k}^2;
\end{align}
\begin{align}
q^i_{n4} 
&= \frac{9}{256}\frac{1}{\pi} S_{n,15}^2 
+ 2\frac{9}{64}\frac{5}{\pi} 
(S_{n,16}^2 + S_{n,17}^2) \nonumber \\
&+ 2\frac{9}{64}\frac{5}{2\pi} 
(S_{n,18}^2 + S_{n,19}^2)
+ 2\frac{9}{64}\frac{35}{\pi} 
(S_{n,20}^2 + S_{n,21}^2)\nonumber \\
&+ 2\frac{9}{256}\frac{35}{2\pi} 
(S_{n,22}^2 + S_{n,23}^2) \equiv \sum_{k=15}^{23}C^\mathrm{3b}_k S_{n,k}^2,
\end{align}
whereby we defined the 3-body coefficients $\{C^\mathrm{3b}_k\}_{k=0}^{23}$.

For 4-body angular descriptor components, we only consider the case of $l_1=l_2=l_3$ and up to $l^\mathrm{4b}_\mathrm{max}=2$. It turns out that $q^i_{n111}=q^i_{n333}=0$. Therefore, there is no difference between $l^\mathrm{4b}_\mathrm{max}=2$ and $l^\mathrm{4b}_\mathrm{max}=3$. Then we only have the case of  $l_1=l_2=l_3=2$,
\begin{align}
q^i_{n222} &=
-\sqrt{\frac{2}{35}} A^i_{n20} A^i_{n20} A^i_{n20} \nonumber \\
& +6\sqrt{\frac{1}{70}} A^i_{n20} A^i_{n21} A^i_{n2(-1)} \nonumber \\
& +6\sqrt{\frac{2}{35}} A^i_{n20} A^i_{n22} A^i_{n2(-2)} \nonumber \\
&-6\sqrt{\frac{3}{35}} A^i_{n21} A^i_{n21} A^i_{n2(-2)}.
\end{align}
The root-rational-fraction package \cite{stone1980cpc} has been used to obtain analytical expressions of the various Wigner $3j$ symbols.
With some algebra, we have 
\begin{align}
q^i_{n222} 
&=-\sqrt{\frac{2}{35}} \left(\frac{1}{4}\sqrt{\frac{5}{\pi}}\right)^3S_{n,3}^3 \nonumber \\
&-6\sqrt{\frac{1}{70}} \left(\frac{1}{4}\sqrt{\frac{5}{\pi}}\right)
\left(\frac{1}{2}\sqrt{\frac{15}{2\pi}}\right)^2
S_{n,3}(S_{n,4}^2+S_{n,5}^2) \nonumber \\
&+6\sqrt{\frac{2}{35}} \left(\frac{1}{4}\sqrt{\frac{5}{\pi}}\right)
\left(\frac{1}{4}\sqrt{\frac{15}{2\pi}}\right)^2
S_{n,3}(S_{n,6}^2+S_{n,7}^2) \nonumber \\
&+6\sqrt{\frac{3}{35}} \left(\frac{1}{2}\sqrt{\frac{15}{2\pi}}\right)^2
\left(\frac{1}{4}\sqrt{\frac{15}{2\pi}}\right)
S_{n,6}(S_{n,5}^2-S_{n,4}^2) \nonumber \\
&-12\sqrt{\frac{3}{35}} \left(\frac{1}{2}\sqrt{\frac{15}{2\pi}}\right)^2
\left(\frac{1}{4}\sqrt{\frac{15}{2\pi}}\right)
S_{n,4} S_{n,5} S_{n,7} \nonumber \\
&\equiv C^\mathrm{4b}_0 S_{n,3}^3 +
C^\mathrm{4b}_1 S_{n,3}(S_{n,4}^2+S_{n,5}^2) \nonumber \\
& + C^\mathrm{4b}_2 S_{n,3}(S_{n,6}^2+S_{n,7}^2) \nonumber \\
& + C^\mathrm{4b}_3 S_{n,6}(S_{n,5}^2-S_{n,4}^2)
+ C^\mathrm{4b}_4 S_{n,4} S_{n,5} S_{n,7},
\end{align}
whereby we defined the 4-body coefficients $\{C^\mathrm{4b}_k\}_{k=0}^4$. 

For 5-body descriptors, we only consider up to $l_1=l_2=l_3=l_4=1$, 
\begin{align}
q^i_{n1111} &=
\frac{7}{15} \left(A^i_{n10}\right)^4  
-\frac{28}{15}\left(A^i_{n10}\right)^2 A^i_{n11} A^i_{n1(-1)}\nonumber \\
&+\frac{28}{15}\left(A^i_{n11}\right)^2 \left(A^i_{n1(-1)}\right)^2\nonumber \\
&= \frac{21}{80\pi^2} S_{n,0}^4 + \frac{21}{40\pi^2} S_{n,0}^2(S_{n,1}^2+S_{n,2}^2) \nonumber \\
& + \frac{21}{80\pi^2} (S_{n,1}^2+S_{n,2}^2)^2\nonumber \\
&\equiv C^\mathrm{5b}_0 S_{n,0}^4 +
C^\mathrm{5b}_1 S_{n,0}^2(S_{n,1}^2+S_{n,2}^2) \nonumber \\
&+ C^\mathrm{5b}_2 (S_{n,1}^2+S_{n,2}^2)^2, 
\end{align}
whereby we defined the 5-body coefficients $\{C^\mathrm{5b}_k\}_{k=0}^2$.

In our implementation, the 3-body coefficients $\{C^\mathrm{3b}_k\}_{k=0}^{23}$, 4-body coefficients $\{C^\mathrm{4b}_k\}_{k=0}^4$, and 5-body coefficients $\{C^\mathrm{5b}_k\}_{k=0}^2$ are pre-computed.
This is crucial for obtaining high computational performance.

We can now enumerate the descriptor vector length.
There are $(n_\mathrm{max}^\mathrm{R}+1)$ radial descriptor components, $(n_\mathrm{max}^\mathrm{A} +1)l_\mathrm{max}^\mathrm{3b}$ 3-body descriptor components, $(n_\mathrm{max}^\mathrm{A} +1)$ 4-body descriptor components, and $(n_\mathrm{max}^\mathrm{A} +1)$ 5-body descriptor components.
Therefore, we have 
\begin{equation}
N_\mathrm{des} = \left(n_\mathrm{max}^\mathrm{R} +1\right) + \left(n_\mathrm{max}^\mathrm{A} +1\right)\left(l_\mathrm{max}^\mathrm{3b} + 2\right)
\end{equation}
in NEP3 if we include both the 4-body and the 5-body descriptor components. 

\subsection{Force, virial, and heat current expressions}

As stressed in Ref.~\onlinecite{fan2017cpc}, we need to derive an explicit expression of the partial force \cite{fan2015prb} for an efficient GPU implementation.

The partial force is 
\begin{align}
\label{equation:partial_force}
\frac{\partial U_i}{\partial \bm{r}_{ij}}
=& \sum_{n=0}^{n_\mathrm{max}^\mathrm{R}} \frac{\partial U_i}{\partial q^i_n}
\frac{\partial q^i_n}{\partial \bm{r}_{ij}}
+ \sum_{n=0}^{n_\mathrm{max}^\mathrm{A}} 
\sum_{l=1}^{l^\mathrm{3b}_\mathrm{max}}
\frac{\partial U_i}{\partial q^i_{nl}}
\frac{\partial q^i_{nl}}{\partial \bm{r}_{ij}} \nonumber \\
+& \sum_{n=0}^{n_\mathrm{max}^\mathrm{A}}
\sum_{l=1}^{l^\mathrm{4b}_\mathrm{max}}\frac{\partial U_i}{\partial q^i_{nlll}}
\frac{\partial q^i_{nlll}}{\partial \bm{r}_{ij}}
+\sum_{n=0}^{n_\mathrm{max}^\mathrm{A}}
\sum_{l=1}^{l^\mathrm{5b}_\mathrm{max}}\frac{\partial U_i}{\partial q^i_{nllll}}
\frac{\partial q^i_{nllll}}{\partial \bm{r}_{ij}}.
\end{align}
Because all the relevant functions here are analytical, it is straightforward to derive explicit expressions for all the partial derivatives in the equation above.

With the partial force available, 
the total force acting on atom $i$ from atom $j$ can be computed as \cite{fan2015prb}
\begin{equation} 
\label{equation:F_ij}
\bm{F}_{ij} = \frac{\partial U_i}{\partial \bm{r}_{ij}} - \frac{\partial U_j}{\partial \bm{r}_{ji}},
\end{equation}
which respects Newton's third law $\bm{F}_{ij}=-\bm{F}_{ji}$.
The total force acting on atom $i$ from all the neighboring atoms is thus
\begin{equation} 
\label{equation:F_i}
\bm{F}_{i} = \sum_{j\neq i} \bm{F}_{ij}.
\end{equation}
From the partial force, one can define the per-atom virial \cite{fan2015prb,Gabourie2021}
\begin{equation} 
\label{equation:virial}
\mathbf{W}_{i} = \sum_{j\neq i} \bm{r}_{ij} \otimes \frac{\partial U_j}{\partial \bm{r}_{ji}}.
\end{equation}
By contracting the per-atom virial above with the velocity $\bm{v}_i$, one can then obtain the per-atom heat current \cite{fan2015prb,Gabourie2021}:
\begin{align} 
\label{equation:Ji}
\bm{J}_{i} 
&=  \mathbf{W}_{i} \cdot \bm{v}_i
= \sum_{j\neq i} \left(\bm{r}_{ij} \otimes 
\frac{\partial U_j}{\partial \bm{r}_{ji}}\right) \cdot \bm{v}_i \nonumber \\
&= \sum_{j\neq i} \bm{r}_{ij} 
\left(\frac{\partial U_j}{\partial \bm{r}_{ji}} \cdot \bm{v}_i \right).
\end{align}
The total heat current in the system is the sum of the per-atom contributions:
\begin{equation}
\label{equation:J-1}
\bm{J} = \sum_i \bm{J}_{i} 
    = \sum_i \sum_{j\neq i} \bm{r}_{ij} 
\left(\frac{\partial U_j}{\partial \bm{r}_{ji}} \cdot \bm{v}_i \right).
\end{equation}
By an exchange of dummy indices, we can also write Eq.~\eqref{equation:J-1} as
\begin{equation}
\label{equation:J-2}
\bm{J} = -\sum_i \sum_{j\neq i} \bm{r}_{ij} 
\left(\frac{\partial U_i}{\partial \bm{r}_{ij}} \cdot \bm{v}_j \right).
\end{equation}
Both Eqs.~\eqref{equation:J-1} and \eqref{equation:J-2} can be used in the Green-Kubo method for thermal conductivity calculations, but Eq.~\eqref{equation:J-1} is a more convenient form for the \gls{hnemd} method and the related spectral decomposition method, as it does not involved the velocities $\bm{v}_j$ of the neighboring atoms $j$ for a given atom $i$.

The heat current expressions above apply to all the interatomic potentials implemented in \textsc{gpumd}.
In all the cases, the validity of the heat current expressions has been numerically confirmed in terms of energy conservation \cite{gill-Comeau2015prb, fan2017prb, xu2018msmse, Brorsson2022ats}.
Equation~\eqref{equation:J-2} has been recently used in an on-the-fly \gls{mlp} \cite{verdi2021npj}.
Since the only assumptions for the derivations \cite{fan2015prb} of these expressions are the locality properties of the interatomic potentials, our heat current expressions generally apply to the multi-body cluster potentials as considered in Ref.~\onlinecite{Boone2019jctc}, as we show in Appendix~\ref{section:heat_current}.

\subsection{Loss function and training algorithm}

We use the \gls{snes} \cite{Schaul2011} to optimize the free parameters in NEP.
We denote a set of parameters as a vector $\mathbf{z}$, whose dimension is the total number of parameters $N_\mathrm{par}$. In NEP1,
\begin{equation}
N_\mathrm{par} = (N_\mathrm{des}+2) N_\mathrm{neu} + 1, 
\end{equation}
which is the same for both single-component and multi-component systems.
In NEP2, we added trainable parameters to the descriptor for multi-component systems and we have 
\begin{equation}
N_\mathrm{par} = (N_\mathrm{des}+2) N_\mathrm{neu} + 1 + N_\mathrm{typ}^2(n^\mathrm{R}_\mathrm{max}+n^\mathrm{A}_\mathrm{max}+2), 
\end{equation}
if $N_\mathrm{typ}>1$, and the same $N_\mathrm{par}$ as in NEP1 if $N_\mathrm{typ}=1$.
In NEP3, the number of trainable parameters in the descriptor is increased for both single and multi-component systems and we have
\begin{align}
N_\mathrm{par} 
& = (N_\mathrm{des}+2) N_\mathrm{neu} + 1  \nonumber \\
& + N_\mathrm{typ}^2(n^\mathrm{R}_\mathrm{max}+1)(N_\mathrm{bas}^\mathrm{R}+1) \nonumber \\
& + N_\mathrm{typ}^2(n^\mathrm{A}_\mathrm{max}+1)(N_\mathrm{bas}^\mathrm{A}+1).
\end{align}
We can formally express the loss function as a function of the free parameters,
\begin{equation}
L = L(\mathbf{z}),
\end{equation}
and express the training process as a real-valued optimization problem:
\begin{equation}
\mathbf{z}^{\ast} = \arg\min_{\mathbf{z}} L(\mathbf{z}),
\end{equation}
where $\mathbf{z}^{\ast}$ is an optimal set of parameters.

The total loss function is defined as a weighted sum of all these individual loss functions:
\begin{align}
\label{equation:loss}
L(\mathbf{z}) 
&= \lambda_\mathrm{e} \left( 
\frac{1}{N_\mathrm{str}}\sum_{n=1}^{N_\mathrm{str}} \left( U^\mathrm{NEP}(n,\mathbf{z}) - U^\mathrm{tar}(n)\right)^2
\right)^{1/2} \nonumber \\
&+  \lambda_\mathrm{f} \left( 
\frac{1}{3N}
\sum_{i=1}^{N} \left( \bm{F}_i^\mathrm{NEP}(\mathbf{z}) - \bm{F}_i^\mathrm{tar}\right)^2
\right)^{1/2} \nonumber \\
&+  \lambda_\mathrm{v} \left( 
\frac{1}{6N_\mathrm{str}}
\sum_{n=1}^{N_\mathrm{str}} \sum_{\mu\nu} \left( W_{\mu\nu}^\mathrm{NEP}(n,\mathbf{z}) - W_{\mu\nu}^\mathrm{tar}(n)\right)^2
\right)^{1/2} \nonumber \\
&+  \lambda_1 \frac{1}{N_\mathrm{par}} \sum_{n=1}^{N_\mathrm{par}} |z_n| \nonumber \\
&+  \lambda_2 \left(\frac{1}{N_\mathrm{par}} \sum_{n=1}^{N_\mathrm{par}} z_n^2\right)^{1/2},
\end{align}
where $N_\mathrm{str}$ is the number of structures in the training data set (if using a full batch) or the number of structures in a mini-batch and $N$ is the total number of atoms in these structures. $U^\mathrm{NEP}(n,\mathbf{z})$ and $W_{\mu\nu}^\mathrm{NEP}(n,\mathbf{z})$ are per-atom energy and virial tensor predicted by the NEP with parameters $\mathbf{z}$ for the $n^{\rm th}$ structure, and $\bm{F}_i^\mathrm{NEP}(\mathbf{z})$ is the predicted force for the $i^{\rm th}$ atom.
$U^\mathrm{tar}(n)$, $W_{\mu\nu}^\mathrm{tar}(n)$ and $\bm{F}_i^\mathrm{tar}$ are the corresponding target values.
That is, the loss functions for energy, force, and virial are defined as their \glspl{rmse} between the current NEP predictions and the target values. 
The last two terms represent $\ell_1$ and $\ell_2$ regularization.
The weights $\lambda_\mathrm{e}$, $\lambda_\mathrm{f}$, $\lambda_\mathrm{v}$, $\lambda_1$, and $\lambda_2$ are tunable hyper-parameters.
When calculating the loss function, we use the following units: \si{\electronvolt\per\atom} for energy and virial and \si{\electronvolt\per\angstrom} for force components.

The \gls{snes} \cite{Schaul2011} we use for optimizing Eq.~\eqref{equation:loss} is a principled approach to real-valued evolutionary optimization by following the natural gradient of the loss function to update a search distribution (a mean value and a variance for each trainable parameter) for a population of solutions.
It is a derivative-free black-box optimizer and thus does not require the loss function to have any analytical properties.
An explicit workflow of the training algorithm has been presented in Ref.~\onlinecite{fan2021neuroevolution}.

\begin{figure*}
\centering
\includegraphics[width=0.9\linewidth]{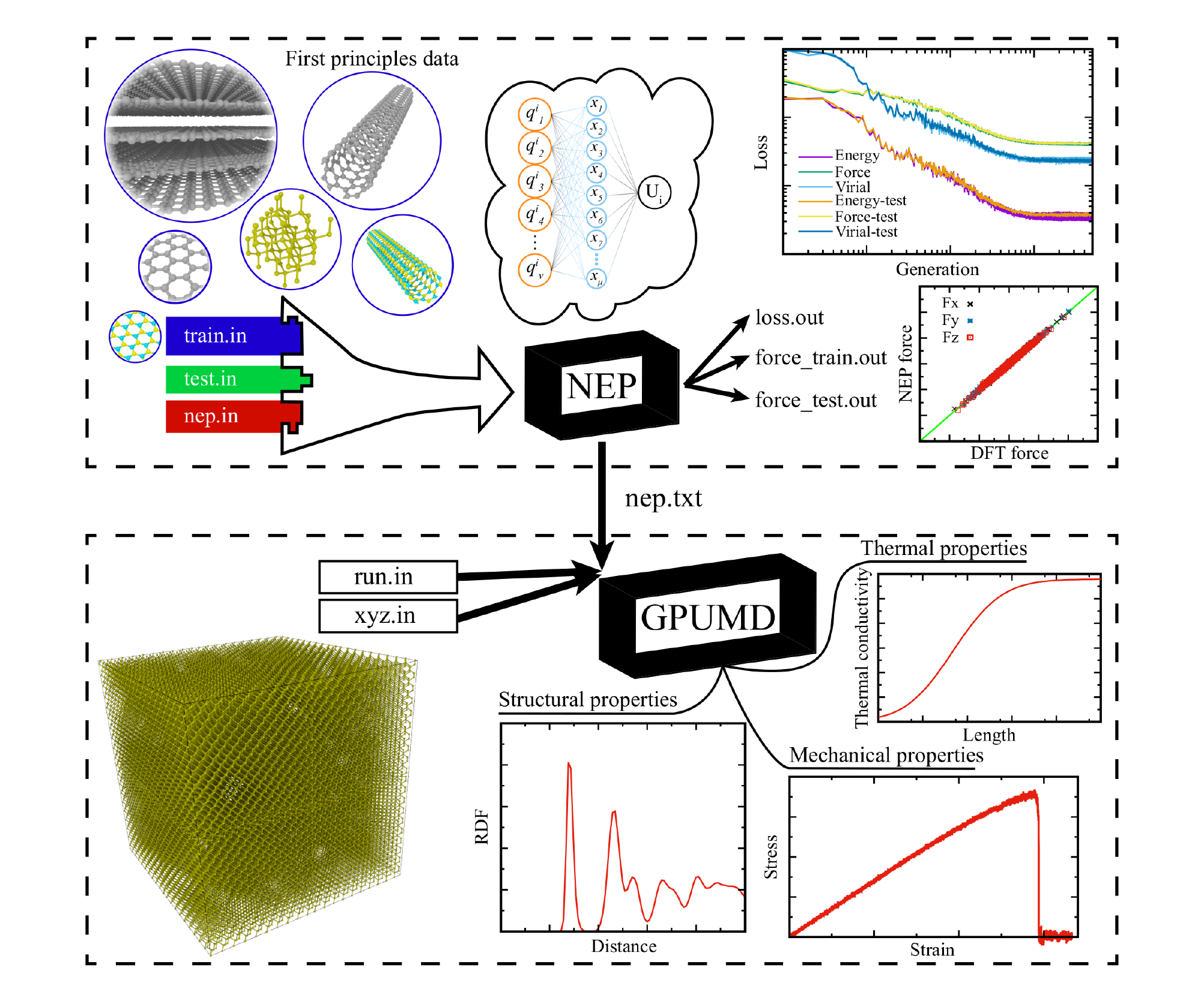}
\caption{
    The \textsc{gpumd} package includes two executables, \texttt{gpumd} and \texttt{nep}, which are represented by the two black boxes.
    The \texttt{nep} program can be used for training NEP models and the \texttt{gpumd} program can be used to perform atomistic simulations using the trained potentials.
    See text for details.
}
\label{fig:workflow}
\end{figure*}

\subsection{GPU implementation}

The NEP approach is implemented in the open-source \textsc{gpumd} package, which is a general-purpose \gls{md} simulation package fully implemented on GPUs.
It currently supports only Nvidia GPUs and the programming language is CUDA C++.
In this section as well as Appendix~\ref{section:algorithms}, we present the detailed algorithms for our CUDA implementation of the NEP approach.

Similar to many other \glspl{mlp}, NEP is a  many-body potential and is very similar to empirical many-body potentials such as the \gls{eam} potential \cite{daw1984prb} and the Tersoff potential \cite{tersoff1989prb}.
Specifically, the radial descriptor part resembles an \gls{eam} potential and the angular descriptor part resembles a Tersoff potential.
Therefore, our CUDA implementation of the NEP approach follows the established efficient scheme for empirical many-body potentials \cite{fan2017cpc}. 

The overall strategy of our CUDA implementation of the NEP approach is to use a few CUDA kernels only, which ensures a high degree of parallelism and high arithmetic intensity, both of which are crucial for achieving high performance in CUDA programming.
In all the CUDA kernels, one CUDA thread is assigned to one atom, i.e., there is a one-to-one correspondence between atoms and CUDA threads.
The descriptor vector and the various per-atom quantities, including the site energy, force, and virial for one atom, are all calculated within one CUDA thread.
This also includes the application of the neural network as represented by Eq.~\eqref{equation:Ui}.
Appendix~\ref{section:ann} shows the CUDA \verb"__device__" function (to be called in the first CUDA kernel function as discussed below) evaluating Eq.~\eqref{equation:Ui} as well as $\{\partial U_i/\partial q^i_{\nu}\}$.
One can see that the feedforward neural network used here is an analytical multi-variable scalar function.

The evaluation of NEP related quantities (energy, force, and virial) requires 4 CUDA kernels only:
\begin{enumerate}
\item The first CUDA kernel calculates the whole  descriptor vector and applies the neural network to obtain the site energy and the derivatives of the energy with respect to the descriptor components, see Algorithm~\ref{algo:energy} in Appendix~\ref{section:algorithms}.
\item The second CUDA kernel calculates the force and virial related to the radial descriptor components, see Algorithm~\ref{algo:radial_force} in Appendix~\ref{section:algorithms}.
\item The third CUDA kernel calculates the partial force related to the angular descriptor components, see Algorithm~\ref{algo:angular_partial_force} in Appendix~\ref{section:algorithms}.
\item The fourth CUDA kernel calculates the force and virial related to the angular descriptor components, see Algorithm~\ref{algo:angular_force} in Appendix~\ref{section:algorithms}.
\end{enumerate}

In these CUDA kernels, the inputs and outputs related to the atoms (position, energy, force, and virial) are all in double precision, but the internal calculations within the CUDA kernels are mostly in single precision. 
This is an effective mixed-precision approach which can keep a good balance between accuracy and efficiency that has been adopted in many other GPU-accelerated atomistic simulation packages \cite{Anderson2008jcp, pall2020jcp, thompson2022cpc}. 

\subsection{Training and using  NEPs}
\label{section:using-nep}

\subsubsection{Overview of the GPUMD package}

The \textsc{gpumd} package \cite{fan2013cpc, fan2017cpc} can be used to train NEPs and use them in atomistic simulations.
\textsc{gpumd} can be compiled and used in both Linux and Windows, provided that a CUDA development environment and a CUDA-enabled GPU are available.
After compilation, one obtains the \verb"nep" and \verb"gpumd" executable, which can be used to train and use NEPs, respectively.
Figure~\ref{fig:workflow} provides a schematic overview of the workflow of training and using NEPs. 

The \textsc{gpumd} package started from a minimal CUDA code implementing only simple pairwise potentials and thermal conductivity calculations \cite{fan2013cpc}.
Gradually, empirical many-body potentials were implemented using the unique formalism we proposed \cite{fan2015prb, fan2017cpc}, including the Tersoff \cite{tersoff1989prb} and Tersoff-like \cite{fan2019jpcm} potentials, the Stillinger-Weber potential \cite{stillinger1985prb}, and \gls{eam} potentials \cite{daw1984prb}.
Recently, support was also added for machine-learned force constant potentials  constructed using the \textsc{hiphive} package \cite{eriksson2019hiphive} \cite{Brorsson2022ats}.
The most recent addition are the various versions of NEP as developed in the previous papers \cite{fan2021neuroevolution, fan2022jpcm} as well as the current one.

Apart from supporting the above important interatomic potentials, \textsc{gpumd} also supports many statistical ensembles, including the NVE (microcanonical), NVT (isothermal), and NPT (isothermal-isobaric) ensembles.
For the NVT ensemble, it has options for the Berendsen thermostat \cite{berendsen1984jcp}, the Nos\'{e}-Hoover chain thermostat \cite{nose1984jcp, hoover1985pra, martyna1992jcp}, the Bussi-Donadio-Parrinello thermostat \cite{Bussi2007jcp}, and the Langevin thermostat in different flavors \cite{Bussi2007pre, Leimkuhler2013jcp}.
For the NPT ensemble, \textsc{gpumd} supports the classical Berendsen barostat \cite{berendsen1984jcp} and the recently proposed stochastic cell-rescaling barostat by Bernetti and Bussi \cite{Bernetti2020jcp}.

\textsc{gpumd} has been mainly used to study thermal transport.
It supports the equilibrium \gls{md} method based on the Green-Kubo relation \cite{green1954jcp, kubo1957jpsj}, the non-equilibrium \gls{md} method using two or more thermostats (preferably the Langevin thermostats \cite{li2019jcp}), and the \gls{hnemd} method using a homogeneous driving force \cite{evans1982pla, fan2019prb}.
The thermal transport coefficients, either thermal conductance or thermal conductivity, can be decomposed both spatially and spectrally \cite{fan2017prb, fan2019prb, Gabourie2021}.
\textsc{gpumd} has been used to establish the best practices \cite{li2019jcp, dong2018prb} in \gls{md} simulations of thermal transport.
It has also been used to study the particular thermal transport properties of specific materials, including various two-dimensional (2D) materials \cite{Mortazavi2016carbon, azizi2017carbon, fan2017nl, Mortazavi2017nanoscale, Mortazavi2018carbon, xu2018msmse, dong2018pccp, xu2019prb,gu2019prb, wu2020cms, wu2020nt, wu2021jpcc, wu2021ijhmt, dong2021jap, ying2022ijhmt, sha2022ct}, \gls{vdw} structures based on 2D materials \cite{Rajabpour2018apl, Gabourie2020td, kim2021nature, wu2021acsami, wu2022ijhmt_1, wu2022ijhmt, feng2022ps}, and quasi-one-dimensional materials \cite{dong2020carbon,Barbalinardo2021prb,liang2022mtp}.
There are applications focused on revealing unique phonon transport mechanisms \cite{Isaeva2019nc, fu2020prb, zhang2021prb, wang2021nanoscale, dong2021prb, Lundgren2021prb, li2022ijhmt}.
The high efficiency of \textsc{gpumd} also enabled high-throughput thermal transport simulations that were used as training/testing data for machine learning models of interfacial thermal transport \cite{jin2022ijhmt}. 

Although previous studies using \textsc{gpumd} have been mostly focused on thermal transport properties, \textsc{gpumd} has already been developed into a general-purpose atomistic simulation package.
In \autoref{section:examples} we will showcase a series of typical atomistic simulations using a NEP model for carbon systems trained in this paper. 

\subsubsection{Training a NEP model}
\label{section:train_nep}

To train a NEP model, one needs to prepare three input files: \verb"train.in", \verb"test.in", and \verb"nep.in".
The first two contain the training and the testing data, respectively.
The files \verb"train.in" and \verb"test.in" have the same data format, the only difference being that the data in \verb"train.in" will be used for training and those in \verb"test.in" will be used for testing.
The file \verb"train.in" (\verb"test.in") contains the following data: the number of structures in the training (testing) set, the reference energy, reference virial tensor (optional), and cell metric for each structure, as well as the chemical symbol, position vector, and reference force vector for each atom in each structure.
For the specific data format, we refer the reader to the \textsc{gpumd} manual.

The file \verb"nep.in" contains hyper-parameters that define the NEP model and control the training process.
In this file, one can choose the NEP version (currently NEP2 or NEP3), specify the number of atom types and their chemical symbols, the cutoff distances $r^\mathrm{R}_\mathrm{c}$ and $r^\mathrm{A}_\mathrm{c}$, the radial function parameters ($n^\mathrm{R}_\mathrm{max}$, $n^\mathrm{A}_\mathrm{max}$, $N^\mathrm{R}_\mathrm{bas}$, and $N^\mathrm{A}_\mathrm{bas}$), the angular expansion parameters ($l^\mathrm{3b}_\mathrm{max}$, $l^\mathrm{4b}_\mathrm{max}$, and $l^\mathrm{5b}_\mathrm{max}$), the weights for the different terms in the loss function ($\lambda_\mathrm{e}$, $\lambda_\mathrm{f}$, $\lambda_\mathrm{v}$, $\lambda_1$, and $\lambda_2$), the number of neurons $N_\mathrm{neu}$ in the hidden layer of the neural network, the batch size $N_\mathrm{bat}$ (number of structures in one batch), the population size $N_\mathrm{pop}$, and the number of generations $N_\mathrm{gen}$ in the \gls{snes} training algorithm.
Details concerning the \verb"nep.in" file are presented in the \textsc{gpumd} manual.

During the training process, predicted energy, force, virial values, the various terms of the loss function (both for the training and the testing data sets), the potential file, and a file used for restarting are continuously updated.
Further details concerning the output files of the \verb"nep" executable can be found in the \textsc{gpumd} manual.

\subsubsection{Using a NEP}

The potential file \verb"nep.txt" contains all the information that constitutes a NEP model and can be used directly as an input to the \verb"gpumd" executable for running atomistic simulations.
To use the \verb"gpumd" executable, one needs to prepare another input file, \verb"run.in", which specifies the simulation process.
Several examples are presented in \autoref{section:examples} below.

\section{Performance of NEP models}
\label{section:performance}

\subsection{NEP3 vs NEP2}

In this section, we demonstrate the workflow of using the \verb"nep" executable to train NEPs and show the enhanced accuracy of NEP3 compared to NEP2 due to the improved radial functions.
Here we use the MgAlCu alloy system, which has been studied previously using a \gls{dp} \cite{jiang2021cpb}.
There are 141,409 structures and we used 90\% of this set for training and 10\% for testing.
There are tools in the \textsc{gpumd} package for preparing the required \verb"train.in" and \verb"test.in" input files for the \verb"nep" executable. 

\begin{table}[thb]
\centering
\setlength{\tabcolsep}{2Mm}
\caption{
    \Glspl{rmse} for energies, forces, and virials for the MgAlCu alloy system using NEP2 and NEP3.
}
\label{table:MgAlCu}
\begin{tabular}{lcc}
\hline
\hline
RMSE & NEP2 & NEP3 \\
\hline
Energies (\si{\milli\electronvolt\per\atom}) & 11 & 10 \\
Forces (\si{\milli\electronvolt\per\angstrom}) & 68 & 63 \\
Virials (\si{\milli\electronvolt\per\atom}) & 43 & 41 \\
\hline
\hline
\end{tabular}
\end{table}

The other input file needed for the \verb"nep" executable is \verb"nep.in".
For our test using NEP2, this file reads as follows:
\begin{Verbatim}[frame=single]
version     2
type        3 Cu Al Mg
cutoff      6 3.5
n_max       15 10
l_max       4
neuron      50	
lambda_1    0.05
lambda_2    0.05
lambda_e    1.0
lambda_f    1.0
lambda_v    0.1
batch       1000
population  50
generation  500000
\end{Verbatim}
For our test using NEP3, it reads:
\begin{Verbatim}[frame=single]
version     3
type        3 Cu Al Mg
cutoff      6 3.5
n_max       12 8
basis_size  12 8
l_max       4
neuron      50	
lambda_1    0.05
lambda_2    0.05
lambda_e    1.0
lambda_f    1.0
lambda_v    0.1
batch       1000
population  50
generation  500000
\end{Verbatim}

The only differences between the hyperparameters used for NEP2 and NEP3 are related to the radial functions.
For NEP2, we use $n_\mathrm{max}^\mathrm{R}=15$ and $n_\mathrm{max}^\mathrm{A}=10$.
For NEP3, we use $n_\mathrm{max}^\mathrm{R}=N_\mathrm{bas}^\mathrm{R}=12$ and $n_\mathrm{max}^\mathrm{A}=N_\mathrm{bas}^\mathrm{A}=8$.
The reason for using smaller $n_\mathrm{max}^\mathrm{R}$ and $n_\mathrm{max}^\mathrm{A}$ values in NEP3 compared to NEP2 is that one radial function in NEP3 is a linear combination of a few basis functions, while one radial function in NEP2 is simply one basis function.
Even in this case, NEP3 can achieve a noticeably higher accuracy than NEP2, as shown in \autoref{table:MgAlCu}.
With these parameters, NEP3 still has a computational speed similar to that of NEP2 in \gls{md} simulations, reaching about \SI{1.5e7}{\atomstepsecond} using one Tesla A100 GPU.

\begin{figure}
\centering
\includegraphics[width=\columnwidth]{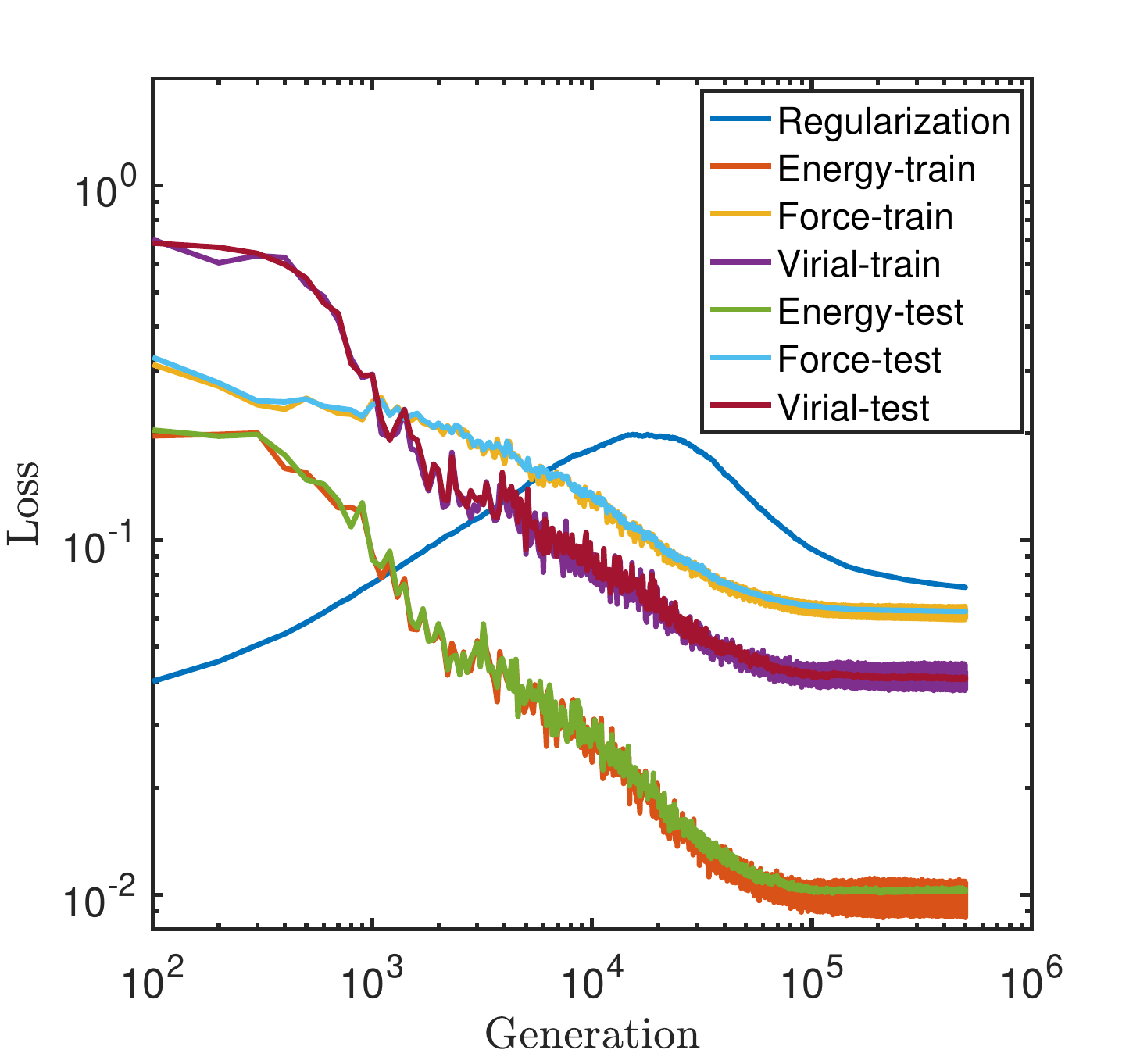}
\caption{
    Evolution of the various terms in the loss function \eqref{equation:loss} with respect to the generation in the \gls{snes} \cite{Schaul2011} training algorithm for the MgAlCu training and test data sets. \cite{jiang2021cpb}
}
\label{fig:loss}
\end{figure}

\begin{figure}
\centering
\includegraphics[width=\columnwidth]{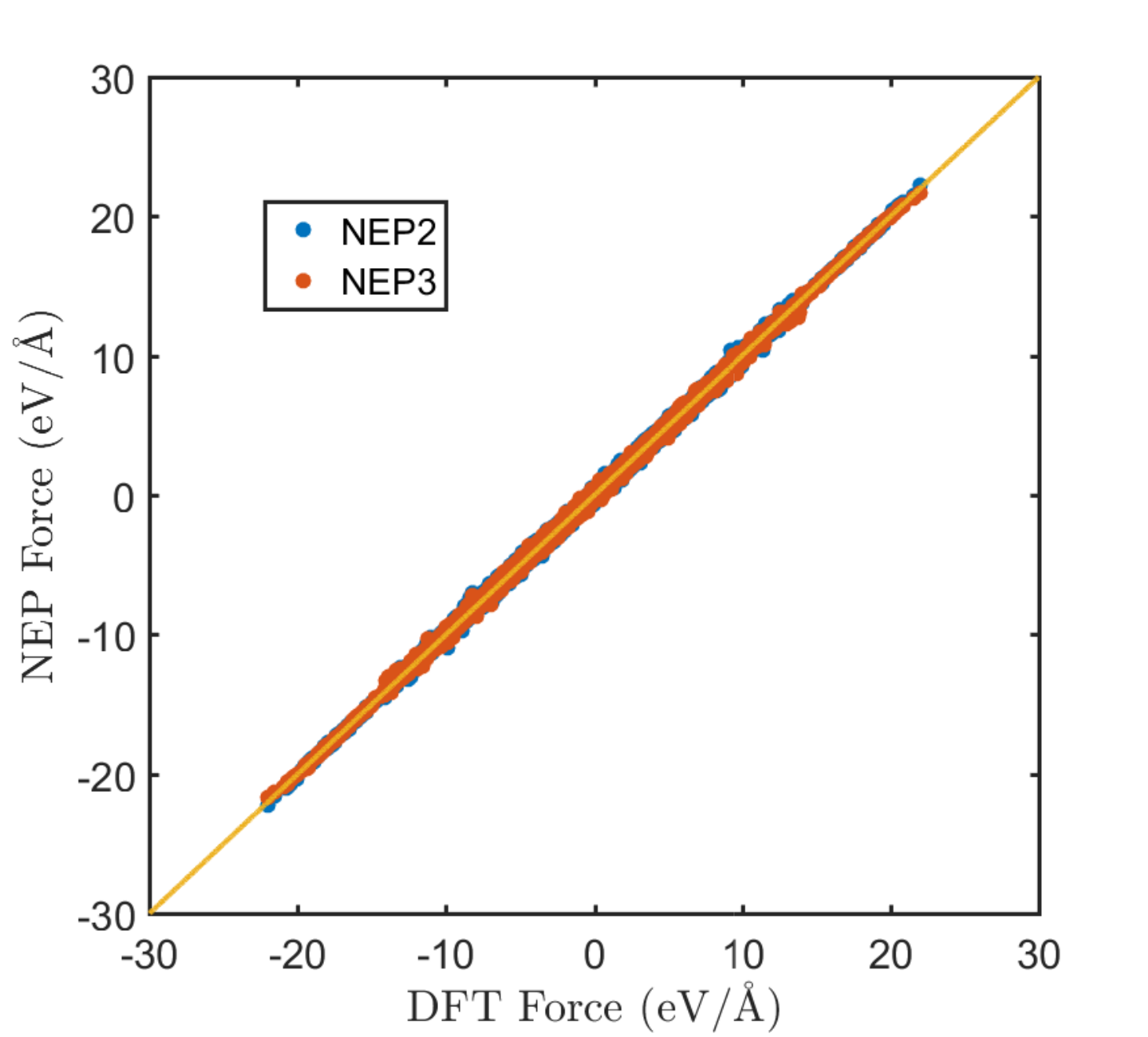}
\caption{
    Forces from NEP2 and NEP3 models against the target \gls{dft} values for the MgAlCu test set.
    The solid line represents the identity function that serves as a guide to the eye.
}
\label{fig:force}
\end{figure}

The \verb"nep" executable produces a number of output files that can be used to examine the training/testing results in detail.
Figure \ref{fig:loss} shows the training and testing \glspl{rmse} as well as the loss function related to the regularization (from the \verb"loss.out" file).
The training \glspl{rmse} exhibits oscillations because of the use of mini-batches (with a batch size of 1,000).
The test \glspl{rmse} closely follow the training \glspl{rmse}, which indicates the very good interpolation capability of NEP.
In other words, there is no sign of over-fitting.
As shown in Ref.~\onlinecite{fan2021neuroevolution}, a proper regularization is crucial to prevent possible over-fitting in NEP models.
Figure~\ref{fig:force} shows the results from the \verb"force_test.out" file.
We can see that both NEP2 and NEP3 achieve a rather high level of accuracy here. 

\subsection{Comparison of NEP3 with other \texorpdfstring{\glspl{mlp}}{MLPs}}
\label{section:compare}

\subsubsection{A general-purpose silicon data set}

We use the general-purpose silicon training data set from Ref.~\onlinecite{bartok2018prx} to test convergence with respect to some hyperparameters and compare the results from an implementation of the \gls{ace} \cite{Dusson2021jcp}.
This data set consists of 2,475 structures, including bulk crystal structures, sp$^2$ bonded structures, dimers, liquid structures, amorphous structures, diamond structures with surfaces or vacancies, and several other defective structures.
Every structure has an energy, but not all the structures have virial data.
For details on the reference \gls{dft} calculations, the reader is referred to Ref.~\onlinecite{bartok2018prx}.

We use the same cutoff distance of \SI{5}{\angstrom} for the radial and angular parts and set $n_\mathrm{max}^\mathrm{R}=n_\mathrm{max}^\mathrm{A} = N_\mathrm{bas}^\mathrm{R}=N_\mathrm{bas}^\mathrm{A}=10$.
We consider using 3-body descriptor components only ($l_\mathrm{max}^\mathrm{3b}=4$, $l_\mathrm{max}^\mathrm{4b}=0$, $l_\mathrm{max}^\mathrm{5b}=0$), using both 3-body and 4-body descriptor components ($l_\mathrm{max}^\mathrm{3b}=4$, $l_\mathrm{max}^\mathrm{4b}=2$, $l_\mathrm{max}^\mathrm{5b}=0$), and using up to 5-body descriptor components ($l_\mathrm{max}^\mathrm{3b}=4$, $l_\mathrm{max}^\mathrm{4b}=2$, $l_\mathrm{max}^\mathrm{5b}=1$).
Other common hyperparameters are as follows: $\lambda_{1}=\lambda_{2}=0.05$, $\lambda_\mathrm{e}=1$, $\lambda_\mathrm{f}=1$, $\lambda_\mathrm{v}=0.1$, $N_\mathrm{neu}=50$, $N_\mathrm{bat}=\rm{full}$, $N_\mathrm{pop}=50$, and $N_\mathrm{gen}=3\times 10^5$.

\begin{figure}
\centering
\includegraphics[width=\columnwidth]{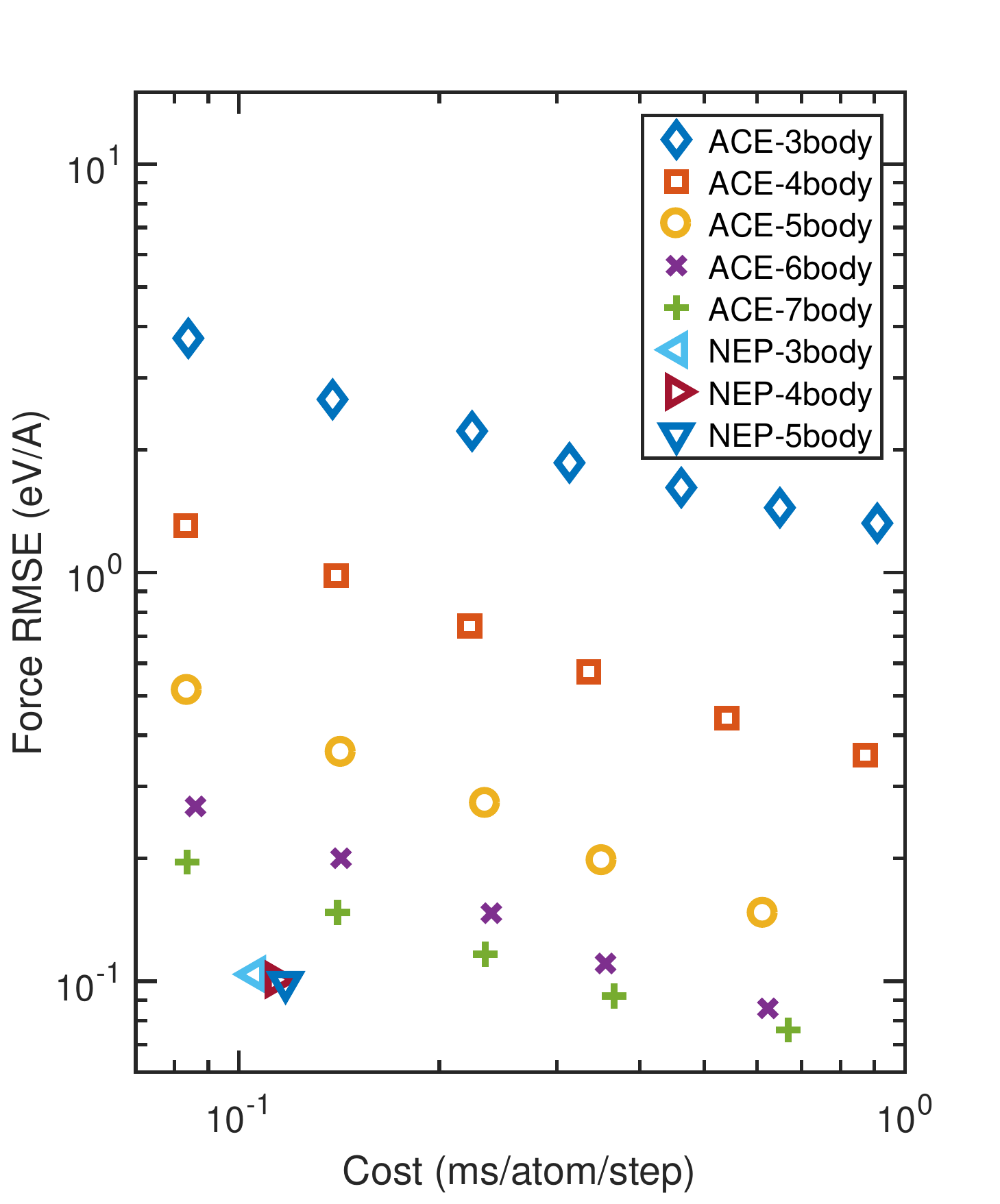}
\caption{
    Force \gls{rmse} against computational cost for NEP models for silicon compared to an implementation of the \gls{ace} approach (the faster recursive approach as in Ref.~\onlinecite{Dusson2021jcp}).
    A serial C++ implementation of the NEP approach has been tested using an Intel i7-8750H CPU @ 2.2 GHz.
    For comparison, the \gls{ace} potential was implemented in Julia and tested using an Intel i7-7820HQ CPU @ 2.9 GHz. \cite{Dusson2021jcp}
}
\label{fig:silicon}
\end{figure}

Figure~\ref{fig:silicon} shows the force \gls{rmse} versus the computational cost of force evaluation for NEP models and previous results for an implementation \cite{Dusson2021jcp} of the \gls{ace} potential \cite{drautz2019prb} based on linear regression.
While the \gls{ace} potential shows a strong dependence of the accuracy on the maximum correlation order of the angular descriptor components, our NEP shows a much weaker dependence, although considering 4-body and 5-body correlations indeed helps to increase the accuracy to some degree.
With 3-body descriptor components only, the \gls{ace} potential has a force \gls{rmse} above \SI{1}{\electronvolt\per\angstrom}, while the 3-body NEP model already achieves an accuracy of about \SI{0.1}{\electronvolt\per\angstrom}.
To achieve the same accuracy as the NEP models, one needs to consider up to 6-body descriptor components in the \gls{ace} approach and the computational cost is a few times larger than that of the NEP models.
The reason for the relative lower cost of the NEP models compared to the \gls{ace} model is probably due to the use of a neural network as the regression method instead of linear regression as used in the \gls{ace} approach.
Using linear regression, the descriptor needs to be rather complete, which can easily lead to more than $10^4$ descriptor components \cite{Dusson2021jcp}, while the descriptor vector lengths range from 55 to 77 in the present NEP models.
With a reduced descriptor length, the completeness \cite{Dusson2021jcp} of the descriptor will be reduced to some degree, but the incompleteness of the descriptor can be (partially) compensated by the neural network and the overall computational cost at a given target accuracy can be lower than using a large number of descriptor components and linear regression.
Indeed, within the framework of $N$-body iterative contraction of equivariants (NICE), nonlinear neural network regression has been shown to be able to achieve higher accuracy than linear regression at least in the limit of large training set size, \cite{Nigam2020jcp} using a descriptor up to the same level of the $N$-body correlation.

\subsubsection{Azobenzene molecule}

Our next example is the largest molecule, azobenzene, in the MD17 data set \cite{Chmiela2017sa}.
This data set has been revised later \cite{christensen2020mlst} to ensure more strict convergence in the \gls{dft} calculations, which is referred to as the revised MD17 data set (rMD17).
Here, we use the first train-test split as reported in rMD17 \cite{christensen2020mlst}, with 1,000 training structures and 1,000 testing structures randomly chosen from a \gls{md} trajectory at \SI{500}{\kelvin}.
There are thus 1,000 target energies and 72,000 target force components in the training data set but there are no target virials.

As we have confirmed that adding 4-body and 5-body descriptor components can lead to higher accuracy, here we use $l^\mathrm{3b}_\mathrm{max}=4$, $l^\mathrm{4b}_\mathrm{max}=2$, $l^\mathrm{5b}_\mathrm{max}=1$ and consider different values of the radial function hyperparameters.
For simplicity, we set $n^\mathrm{R}_\mathrm{max}=N^\mathrm{R}_\mathrm{bas}$ and $n^\mathrm{A}_\mathrm{max}=N^\mathrm{A}_\mathrm{bas}$ and consider the following combinations of parameters: $n^\mathrm{R}_\mathrm{max}=n^\mathrm{A}_\mathrm{max}=(6,4)$, $(9,6)$, $(12,8)$, and $(15,10)$.
The other hyperparameters are: $r_\mathrm{c}^\mathrm{R}=6$ \AA, $r_\mathrm{c}^\mathrm{A}=4$ \AA, $\lambda_{1}=\lambda_{2}=0.02$, $\lambda_\mathrm{e}=1$, $\lambda_\mathrm{f}=1$, $\lambda_\mathrm{v}=0$, $N_\mathrm{neu}=50$, $N_\mathrm{bat}=$ full, $N_\mathrm{pop}=50$, and $N_\mathrm{gen}=10^6$.

\begin{figure}
\centering
\includegraphics[width=\columnwidth]{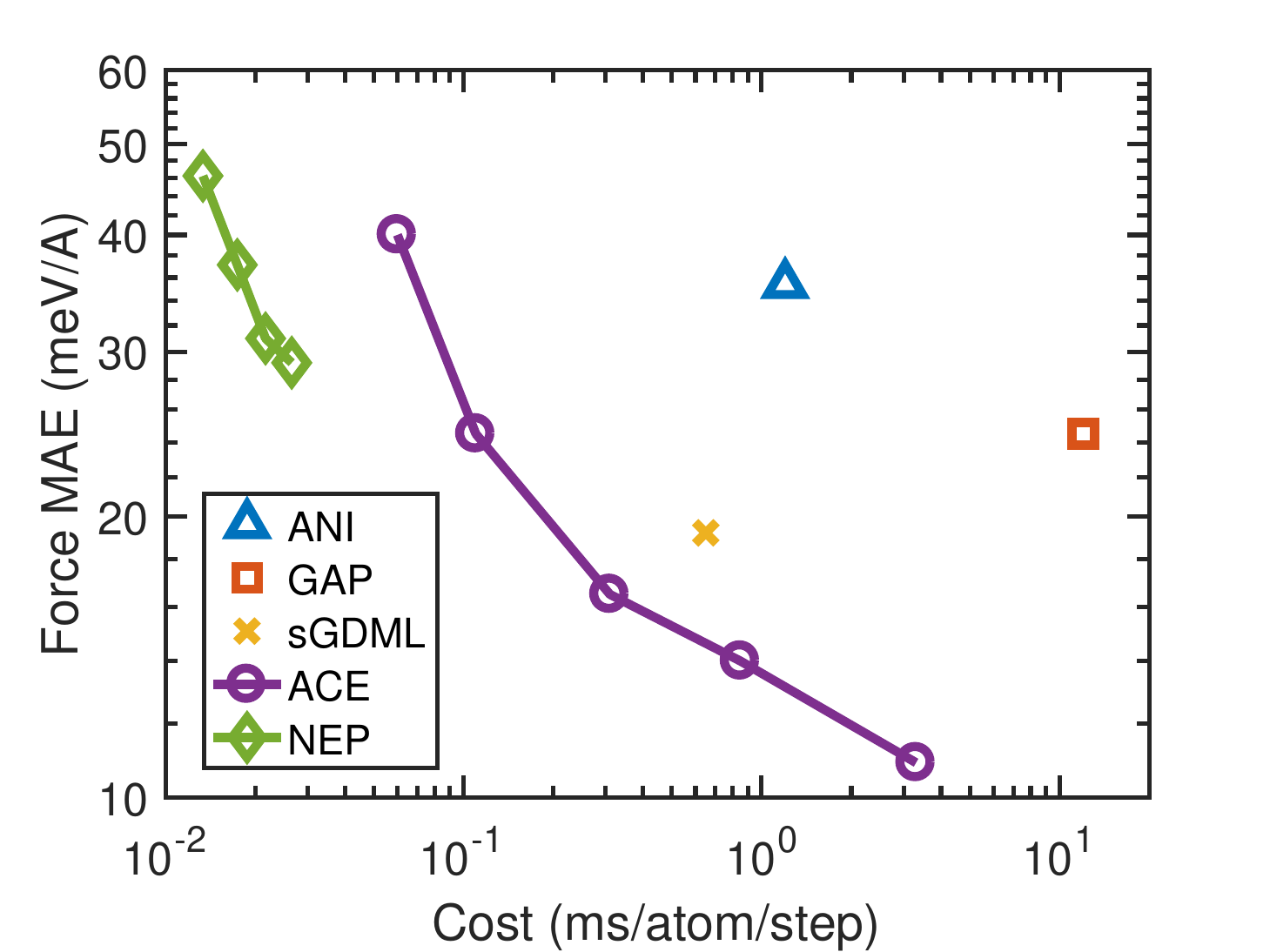}
\caption{
    Force \gls{mae} against computational cost from NEP models and several other \glspl{mlp} for azobenzene as reported in Ref.~\onlinecite{kovacs2021jctc}.
    To be consistent with Ref.~\onlinecite{kovacs2021jctc}, we tested our serial C++ code using an Intel Xeon Gold 5218 @ 2.3 GHz.
}
\label{fig:azobenzene}
\end{figure}

Figure~\ref{fig:azobenzene} shows the force \gls{mae} versus the computational cost of force evaluation for our NEP and some other \glspl{mlp} as reported in Ref.~\onlinecite{kovacs2021jctc}, including ANI \cite{smith2017cs}, \gls{gap} \cite{bartok2010prl}, sGDML \cite{Chmiela2019cpc}, and a linear-regression based \gls{ace} potential \cite{kovacs2021jctc} (similar to but not identical to the one in Ref.~\onlinecite{Dusson2021jcp}).
With increasing $n_\mathrm{max}^\mathrm{R}$ and $n_\mathrm{max}^\mathrm{A}$, both the accuracy and computational cost increase quickly.
With  $n_\mathrm{max}^\mathrm{R}=15$ and $n_\mathrm{max}^\mathrm{A}=10$, the NEP approach achieves an accuracy between ANI and \gls{gap}, but it is more than one order of magnitude faster than ANI and more than two orders of magnitude faster than \gls{gap}.
At this level of accuracy, the NEP models are also several times faster than linear-\gls{ace} potentials, similar to the case of silicon above (\autoref{fig:silicon}).
Similar to the case of of silicon, we propose that the superior cost effectiveness of the NEP  models as compared to the linear-\gls{ace} potentials is due to the much smaller descriptor vector size in NEP, which ranges from 32 to 82 in the NEP models, but from 1,700 to 122,000 in the linear-\gls{ace} potentials for the test cases in Fig.~\ref{fig:azobenzene}.

\subsubsection{Carbon data set}
\label{section:train_carbon}

In this section, we use the training and test data sets for the carbon system with various phases \cite{Deringer2017prb} to compare the NEP approach with other \glspl{mlp} in terms of accuracy, speed, and memory usage.
The \glspl{mlp} to be compared include \gls{dp} \cite{wang2018cpc}, \gls{gap} \cite{bartok2010prl}, \gls{mtp} \cite{Shapeev2016}, and \gls{reann} \cite{zhang2021prl}.
The training data set comprises 4,080 structures in total, including bulk crystal structures, bulk liquid and amorphous structures, amorphous surfaces, and isolated dimer structures.
The testing data set comprises 450 structures similar to those in the training data set, but excluding the dimer ones.
There are 256,628 and 28,337 atoms in the training and testing data sets, respectively.
Each structure has one target energy and each atom has three target force components.
Some structures also have target virials.
For more details on the data sets, see Ref.~\onlinecite{Deringer2017prb}.

\begin{table}[thb]
\centering
\setlength{\tabcolsep}{2Mm}
\caption{
    Performance comparison between NEP models and other \glspl{mlp} for the carbon test set from Ref.~\onlinecite{Deringer2017prb}.
    Energy \gls{rmse} $\Delta E$ and virial \gls{rmse} $\Delta W$ are in units of \si{\milli\electronvolt\per\atom} while the force \gls{rmse} $\Delta F$ is in units of \si{\milli\electronvolt\per\angstrom}.
    Computational speed is measured in \si{\atom\step\per\milli\second}.
    $N_\mathrm{max}$ is the maximum number of atoms that can be simulated using one Tesla V100 GPU for the three GPU-accelerated codes.
    For \gls{gap} \cite{bartok2010prl, Deringer2017prb} and \gls{mtp} \cite{Shapeev2016, Novikov2021}, 72 Intel Xeon-Gold $6240$ CPU cores were used.
    For \gls{dp} \cite{wang2018cpc} after compression, NEP and \gls{reann} \cite{zhang2021prl, zhang2022jcp}, a 32-GB Tesla V100 GPU was used.
}
\label{table:carbon}
\begin{tabular}{l*{6}r}
\hline
\hline
\gls{mlp} &  \multicolumn{1}{c}{$\Delta E$}  & \multicolumn{1}{c}{$\Delta F$} &
\multicolumn{1}{c}{$\Delta W$}  &\multicolumn{1}{c}{Speed} & \multicolumn{1}{c}{$N_\mathrm{max}$} \\
\hline
\gls{gap} & $46$ & $1,100$ & NA & 6.1 & NA \\
\hline
\gls{dp} (se2) & $80$ & $1,100$ & 250 & 290 & $240\times10^3$ \\
\gls{dp} (se2+se3) & $44$ & $800$ & 170  & 150 & $220\times10^3$ \\
\hline
\gls{mtp} ($4$ \AA) & $36$ & $650$ & 180 & 110 & NA \\
\gls{mtp} ($5$ \AA) & $35$ & $630$ & 200 & 61 & NA \\
\gls{mtp} ($6$ \AA) & $35$ & $650$ & 220 & 27 & NA \\
\hline
NEP (\SI{4.2}{\angstrom}) & $42$ & $690$ & 160 & 3,600 & $4,100\times10^3$ \\
NEP (\SI{3.7}{\angstrom}) & $44$ & $700$ & 170 & 4,600 & $5,800\times10^3$ \\
\hline
\gls{reann} (\SI{3}{\angstrom}) & $41$ & $700$ & NA & 280 & $290\times10^3$ \\
\gls{reann} (\SI{4}{\angstrom}) & $31$ & $640$ & NA & 170 & $180\times10^3$ \\
\gls{reann} (\SI{6}{\angstrom}) & $28$ & $670$ & NA & 62 & $64\times10^3$ \\
\hline
\hline
\end{tabular}
\end{table}

In the case of the NEP approach, we used the NEP3 form and considered two sets of hyperparameters.
In the first one, we set $r_\mathrm{c}^\mathrm{R}=\SI{4.2}{\angstrom}$, $r_\mathrm{c}^\mathrm{A}=\SI{3.7}{\angstrom}$, $n_\mathrm{max}^\mathrm{R}=N_\mathrm{bas}^\mathrm{R}=10$, $n_\mathrm{max}^\mathrm{A}=N_\mathrm{bas}^\mathrm{A}=8$, $l^\mathrm{3b}_\mathrm{max}=4$, $l^\mathrm{4b}_\mathrm{max}=2$, $l^\mathrm{5b}_\mathrm{max}=1$, $\lambda_{1}=\lambda_{2}=0.05$, $\lambda_\mathrm{e}=1$, $\lambda_\mathrm{f}=1$, $\lambda_\mathrm{v}=0.1$, $N_\mathrm{neu}=100$, $N_\mathrm{bat}=\rm{full}$, $N_\mathrm{pop}=50$, and $N_\mathrm{gen}=5\times 10^5$.
This model is labelled ``NEP (\SI{4.2}{\angstrom})'' in \autoref{table:carbon}.
In the second one, we make the following changes as compared to the first one: $r_\mathrm{c}^\mathrm{R}=\SI{3.7}{\angstrom}$, $r_\mathrm{c}^\mathrm{A}=\SI{3.2}{\angstrom}$, $l^\mathrm{5b}_\mathrm{max}=0$, and $N_\mathrm{neu}=50$.
This model is labelled ``NEP (\SI{3.7}{\angstrom})'' in \autoref{table:carbon}.

For \gls{dp}, we used the \textsc{DeePMD-kit} package \cite{wang2018cpc} and the smooth edition \cite{zhang2018endtoend}.
We trained two versions of \gls{dp}, one using the \verb"se_a" descriptor (with a cutoff of \SI{6}{\angstrom}) only, and the other using a combination of \verb"se_e2_a" (with a cutoff of \SI{6}{\angstrom} and \verb"se_e3" (with a cutoff of \SI{3.8}{\angstrom}).
These two versions are labelled ``\gls{dp} (se2)'' and ``\gls{dp}(se2+se3)'' in \autoref{table:carbon}, respectively.
The size of the embedding net is $(25,50,100)$ for the \verb"se_a" and \verb"se_e2_a" descriptors and $(20,40,80)$ for the \verb"se_e3" descriptor, and the size of the fitting net is $(240,240,240)$.
The learning rate decreases exponentially from $10^{-3}$ to $10^{-8}$.
The weighting parameters for energy, force, and virial have a starting value of 0.02, 1,000, and 0.01, respectively, which are linearly changed to 1, 1, and 0.1 during the training process.
The number of training steps is $10^7$, which is sufficiently large.

For \gls{gap} \cite{bartok2010prl,Deringer2017prb}, we directly took the results from Ref.~\onlinecite{Deringer2017prb}.
The $n_\mathrm{max}$ and $l_\mathrm{max}$ for the \gls{soap} descriptor were both set to $8$ \cite{Deringer2017prb}.
There were also separate low-dimensional 2-body and 3-body components in this \gls{gap} \cite{Deringer2017prb}.

For \gls{mtp} \cite{Shapeev2016}, we used the \textsc{mlip} package \cite{Novikov2021}.
The descriptor level of the \gls{mtp} is set to $22$.
We considered three  cutoff distances: \SI{4}{\angstrom}, \SI{5}{\angstrom}, and \SI{6}{\angstrom}, labelled ``\gls{mtp} (\SI{4}{\angstrom})'',  ``\gls{mtp} (\SI{5}{\angstrom})'', and ``\gls{mtp} (\SI{6}{\angstrom})'', respectively, in \autoref{table:carbon}.

For \gls{reann} \cite{zhang2021prl}, we used the \textsc{reann} package \cite{zhang2022jcp}.
We considered three cutoff distances: \SI{3}{\angstrom}, \SI{4}{\angstrom}, and \SI{6}{\angstrom}, labelled ``\gls{reann} (\SI{3}{\angstrom})'',  ``\gls{reann} (\SI{4}{\angstrom})'', and ``\gls{reann} (\SI{6}{\angstrom})'', respectively, in \autoref{table:carbon}.
The weighting parameter for energy is kept at 1 and that for force is decreased from 10 to 0.5 during the training process.
A batch-size of 32 is used (we have tried to use a larger batch-size and it turned out to exceed the memory limit of a 32-GB V100).
The sizes of the neural network for the atom energy and the orbital coefficients are both $(64,64)$.
The learning rate decreases exponentially from $10^{-3}$ to $10^{-7}$. The number of training epochs is $10^4$.

For all \glspl{mlp}, we list the \glspl{rmse} for energies, forces, and virials (calculated from 6 independent components) in \autoref{table:carbon}.
The \gls{mtp} and \gls{reann} models show the best accuracy in energies and forces, and \gls{gap} and \gls{dp} the worst.
The accuracy of the NEP models is close to those of \gls{mtp} and \gls{reann}.
The virial \gls{rmse} is missing for \gls{gap} as no predicted virial data have been provided \cite{Deringer2017prb}.
It is also missing for \gls{reann} because the virial has not been formulated in this \gls{mlp}.
For the \glspl{mlp} with available virial data, the NEP models achieve the highest accuracy.
Therefore, we can say that the NEP models at least have an above-average accuracy for this carbon data set.

\begin{figure}
\centering
\includegraphics[width=\columnwidth]{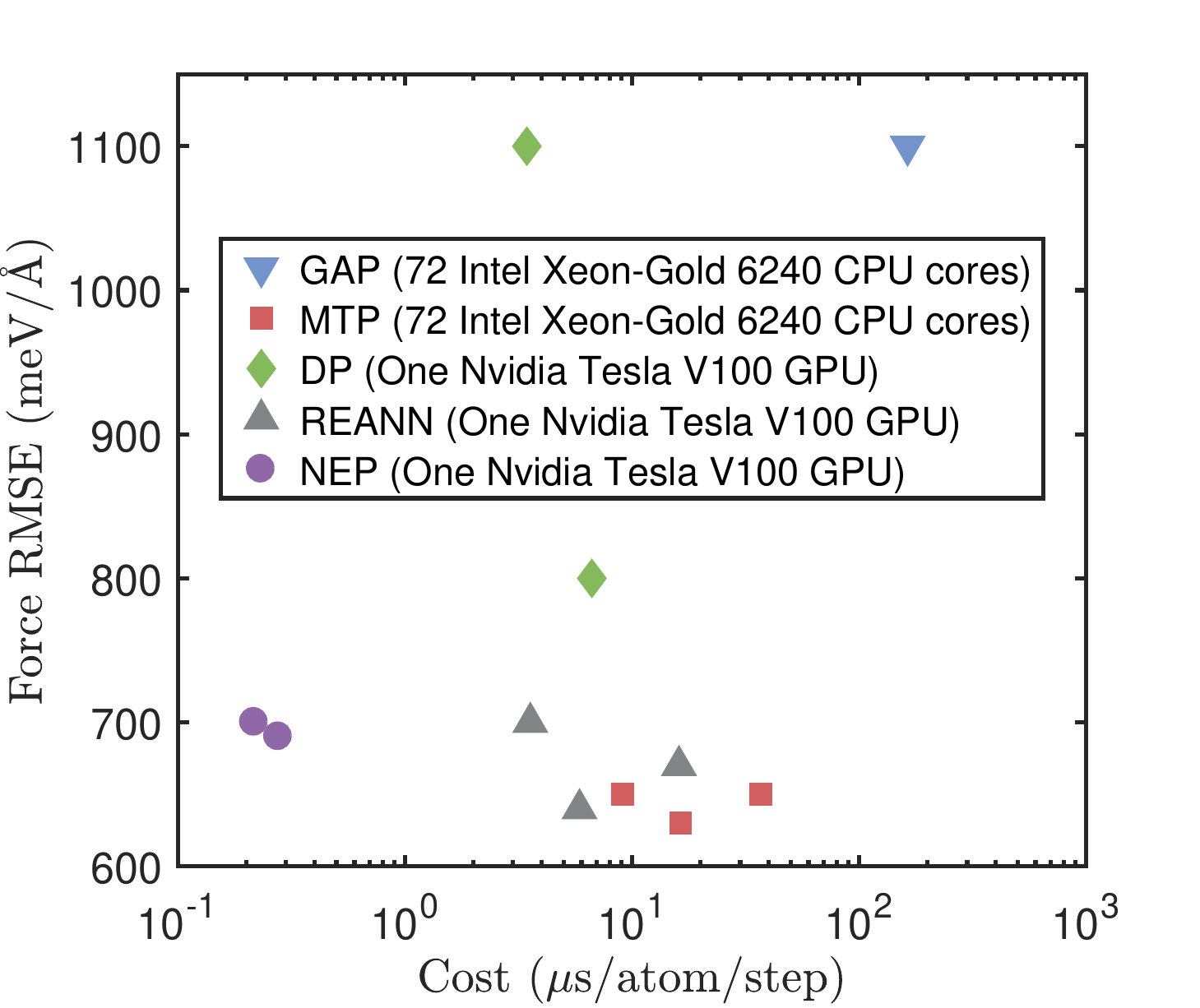}
\caption{
    Force \gls{rmse} against computational cost from NEP models and other \glspl{mlp} for the carbon data set from Ref.~\onlinecite{Deringer2017prb}.
}
\label{fig:carbon}
\end{figure}

With the accuracy comparison results in mind, we next compare the computational performance in realistic atomistic simulations.
Here, we run \gls{md} simulations for a cubic cell of diamond in the isothermal ensemble at \SI{300}{\kelvin} for 100 steps and output some basic thermodynamic properties every 10 steps.
Based on the \gls{md} simulations, we measure the computational speed as the product of the number of atoms and the number of steps divided by the total wall time used.
Three of the \glspl{mlp} (\gls{dp}, \gls{reann}, NEP) have been implemented on GPUs and we thus use an Nvidia Tesla V100 GPU for the test.
For the other two \glspl{mlp} for which there are only CPU versions available (\gls{gap} and \gls{mtp}) we use 72 Intel Xeon-Gold 6240 cores.
The CPU and GPU resources might have unequal financial costs, but one can make suitable conversions of the results presented here to other computational environments.
For the CPU-based \glspl{mlp}, \gls{mtp} shows much higher computational speed than \gls{gap}, which is consistent with previous tests \cite{Shapeev2016}.
For the GPU-based \glspl{mlp}, \gls{dp} (after model compression) and \gls{reann} have comparable speed, while the NEP models are more than one order of magnitude faster. Figure~\ref{fig:carbon} shows that the NEP models substantially lower the Pareto front of accuracy-versus-cost that can be achieved by the other \glspl{mlp}.

Interestingly, the GPU memory usage seems to be correlated to the computational speed: the maximum number of atoms $N_\mathrm{max}$ that can be simulated using one Tesla V100 GPU is roughly proportional to the computational speed.
This comparison highlights the superior computational performance of the NEP approach as implemented in \textsc{gpumd} in terms of both computational speed and memory efficiency, which is crucial for tackling challenging applications that require large-scale and long-time atomistic simulations.

\section{Active learning based on the latent space}
\label{section:active-learning}

Apart from regression accuracy and \gls{md} speed, training data preparation is an important aspect of \glspl{mlp}.
Because quantum-mechanical calculations are usually time consuming, it is desirable to construct a minimal training data set for a given application.
One strategy for achieving this is to use \gls{al}, which uses query criteria to determine whether or not a new training sample should be included into an existing training set to improve the model accuracy and generalization capability.
Many \gls{al} schemes have been proposed for \glspl{mlp}, including the ensemble (or query-by-committee) method  \cite{Gastegger2017cs, Smith2018jcp, zhang2019prm}, the dropout method \cite{wen2020npj}, methods based on feature-space distance measuring \cite{Schran2020jctc} and entropy maximization in the descriptor space \cite{KarPer20}, and methods based on optimal design \cite{Podryabinkin2017cms, Zaverkin2021mlst}.
Here, we propose an \gls{al} scheme based on the latent space of a pre-trained \gls{nep} model.
This \gls{al} scheme has been inspired by the work of Janet \textit{et al.} \cite{Janet2019}, who have shown that distance in the latent space provides a good quantitative uncertainty metric to be used in an \gls{al} scheme.

There is no unique definition of the latent space.
Here, we define it as an $N_\mathrm{neu}$-dimensional space spanned by the vectors whose components are the product of the states of the hidden-layer neurons and the connection weights between them and the output layer.
To compute the latent-space vector for a structure, one must train a \gls{nep} model first, but this can be achieved by using a small initial training data set.
Then one can use the pre-trained \gls{nep} model to compute the latent-space vectors for many structures, either those in the training data set or new ones that have no target values (energy, forces, and virials) yet.
This allows one to generate target values (via quantum-mechanical calculations) for a number of structures that have relatively large distances to existing points in the latent space.
This procedure can be iterated by updating the training data set and the \gls{nep} model in alternating fashion.
During this process, the existing \gls{nep} model can be used to create the new structures to be examined, using various sampling techniques in atomistic simulations.

\begin{figure}
\centering
\includegraphics[width=\columnwidth]{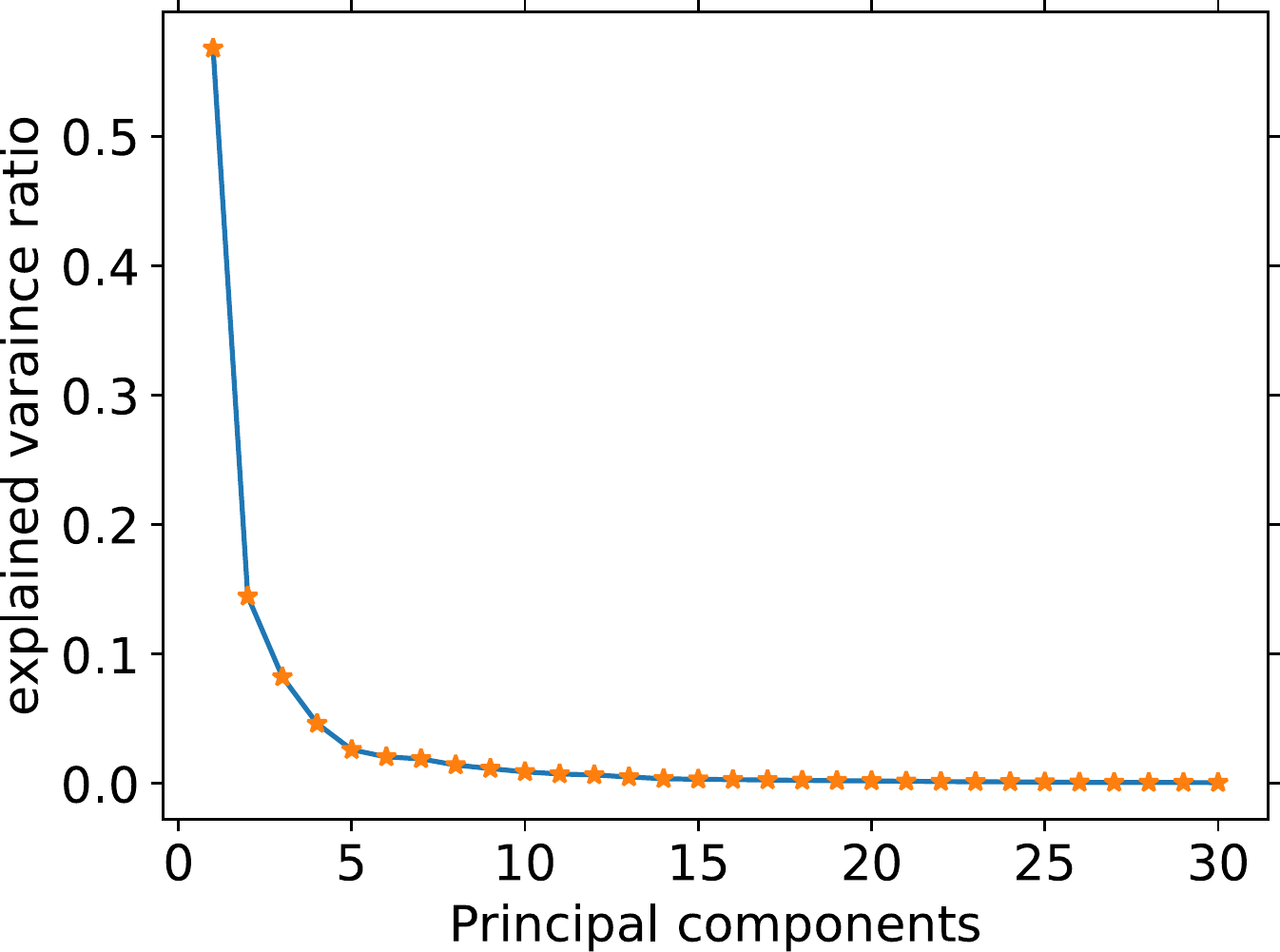}
\caption{
    Normalized explained variance ratio of the first 30 principal components calculated from the 4,080 structures in the original training data set \cite{Deringer2017prb} based on the initial \gls{nep} model trained using 200 structures.
}
\label{fig:pca_variance}
\end{figure}

\begin{figure}
\centering
\includegraphics[width=\columnwidth]{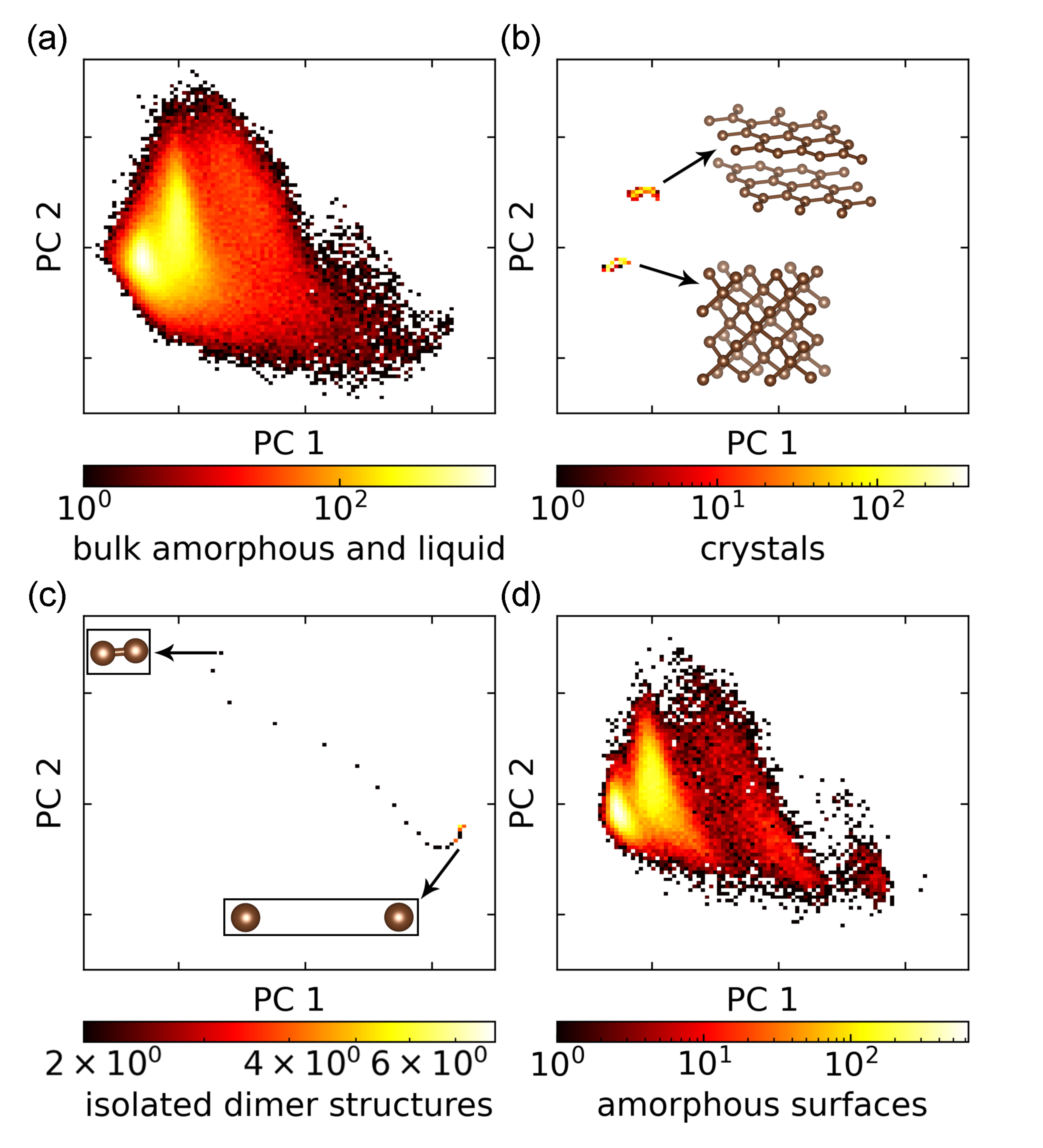}
\caption{
    Distribution of the 4,080 carbon structures in the full training data set \cite{Deringer2017prb} in the 2D principal component (PC) space (spanned by PC 1 and PC 2) as reduced from the latent space that was constructed using the initial NEP model trained using 200 structures.
    (a) bulk amorphous/liquid structures, (b) crystals including sp$^2$ graphite and sp$^3$ diamond structures, (c) dimers, and (d) surface amorphous structures.
    The color bar represents the density of structures in the 2D PC space.
}
\label{fig:latent}
\end{figure}

We take the carbon data set as a concrete example to illustrate the idea outlined above.
To this end, suppose we only have 200 structures randomly selected out of the 4,080 ones in the original training data set.
We first train an initial \gls{nep} using these 200 structures, adopting the same hyperparameters as used for the \gls{nep} (\SI{4.2}{\angstrom}) model in \autoref{table:carbon}.
Assuming that we have obtained the remaining structures as in the original training data set by various means, we then compute the latent space vectors for all the 4,080 structures.
One can define a distance in the high-dimensional latent space, but it turns out that the high dimension can be effectively reduced using \gls{pc} analysis.
The explained variance ratios of the first 30 \glspl{pc} are shown in \autoref{fig:pca_variance}.
The first two leading \glspl{pc} contribute more than 70\% to the total dimensions, allowing us to visualize the distribution of structures in a 2D \gls{pc} space, as shown in \autoref{fig:latent}.
It can be seen that both dimers and crystals (with or without defects) occupy a small area in the \gls{pc} space.
On the other hand, both bulk and surface amorphous/liquid structures occupy large areas that are almost overlapping.
Furthermore, the high density part in \autoref{fig:latent}a indicates that there are relatively more bulk amorphous and liquid structures in the full training data set. 

\begin{figure}
\centering
\includegraphics[width=\columnwidth]{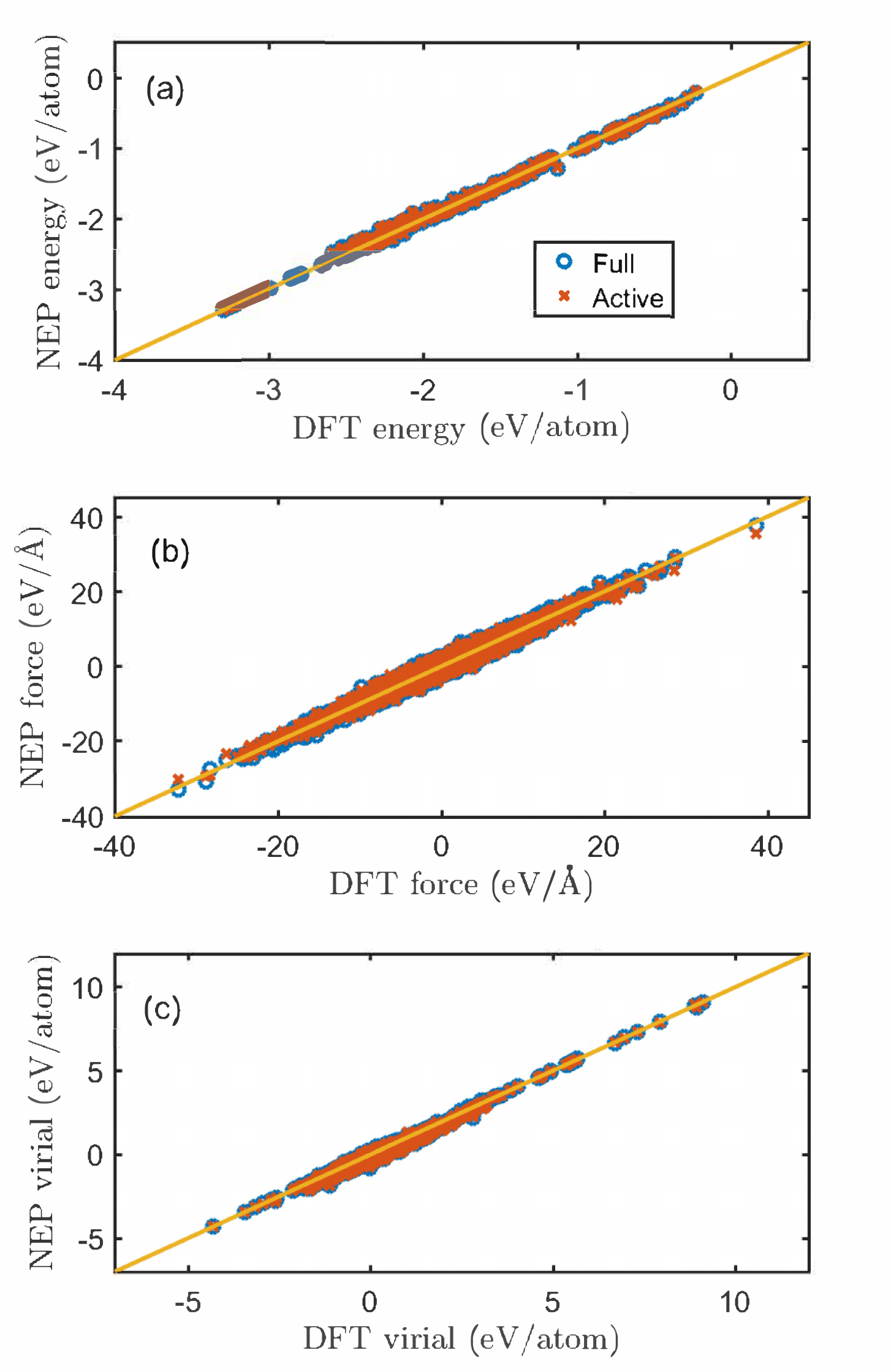}
\caption{
    (a) Energy, (b) force, and (c) virial values from the \gls{nep} models for carbon constructed using the full training data set (4,080 structures, labelled ``Full'') and the training data set constructed based on active learning (786 structures, labelled ``Active''), in comparison to the \gls{dft} reference data for the test data set (450 structures).\cite{Deringer2017prb}
}
\label{fig:active_vs_full}
\end{figure}

Based on these observations, we construct a new training data set which includes all the dimers (30 in total) and crystals (356 in total), and 400 bulk amorphous/liquid structures.
Using these 786 structures, we train a new NEP model using the same hyperparameters as for the NEP (\SI{4.2}{\angstrom}) model in \autoref{section:train_carbon}).
The energy, force, and virial \glspl{rmse} from this NEP model are \SI{45}{\milli\electronvolt\per\atom}, \SI{700}{\milli\electronvolt\per\angstrom}, and \SI{190}{\milli\electronvolt\per\atom}, respectively, for the same test data set as used in \autoref{section:train_carbon}, which are very close to those for the NEP model trained using the full training data set.
Figure~\ref{fig:active_vs_full} shows that the \gls{al}-based NEP model indeed performs very well in the various predicted values.
The force \gls{rmse} (\SI{660}{\milli\electronvolt\per\angstrom}) for the \gls{al}-based NEP model is only slightly higher than that for the NEP model trained using the full training data set (\SI{650}{\milli\electronvolt\per\angstrom}).
This is a notable result since we have not included a single surface amorphous structure into the training data set for the \gls{al}-based NEP model.
This shows that the distance in the latent space (and the reduced \gls{pc} space) indeed provides a reliable metric for selecting new samples for the construction of accurate and transferable \glspl{mlp}.
The present results also indicate that the NEP approach is quite data efficient, which we attribute to the relatively simple neural-network model and the inclusion of regularization terms in the loss function.
As a further demonstration of the reliability of the \gls{al}-based NEP model, we show in \autoref{section:examples} below that it performs equally well as the NEP model trained against the full training data set in an \gls{md} simulation covering a large range of temperatures.

\section{Examples for applications of NEP models}
\label{section:examples}

In this section, we demonstrate the application of NEP models in atomistic simulations.
To this end, we employ the NEP (\SI{4.2}{\angstrom}) model from \autoref{table:carbon}, if not stated otherwise.

\subsection{Lattice constant}

We begin with a simple static calculation and determine the zero-temperature lattice constant of diamond by calculating a cohesive energy curve.
The \verb"run.in" input file reads:
\begin{Verbatim}[frame=single]
potential  potentials/nep/C_2022_NEP3.txt 0
compute_cohesive  0.98 1.03 51
\end{Verbatim}
The \verb"potential" keyword specifies the NEP model to be used and the \verb"compure_cohesive" keyword is used to invoke the cohesive energy calculation.
The lattice constant thus obtained is \SI{3.530}{\angstrom}, which is very close to the \gls{dft} reference value obtained using the \gls{lda}, see \autoref{table:lattice_constant}.

\begin{table}[thb]
\centering
\setlength{\tabcolsep}{2Mm}
\caption{
    Structural and elastic properties of diamond from the NEP (\SI{4.2}{\angstrom}) model in comparison to \gls{gap} and \gls{dft}-\gls{lda} results. \cite{Deringer2017prb}
}
\label{table:lattice_constant}
\begin{tabular}{l*{3}{d}}
\hline
\hline
& \multicolumn{1}{c}{\gls{dft}-\gls{lda}}
& \multicolumn{1}{c}{\gls{gap}}
& \multicolumn{1}{c}{NEP (\SI{4.2}{\angstrom})} \\
\hline
$a$ (\si{\angstrom})         & 3.532 & 3.539 & 3.530  \\
$C_{11}$ (\si{\giga\pascal}) & 1,101 & 1,090 & 1,134  \\
$C_{12}$ (\si{\giga\pascal}) &   148 &   112 &   153  \\
$C_{44}$ (\si{\giga\pascal}) &   592 &   594 &   605  \\
\hline 
\hline
\end{tabular}
\end{table}

\subsection{Elastic constants}

\begin{figure}
\centering
\includegraphics[width=\columnwidth]{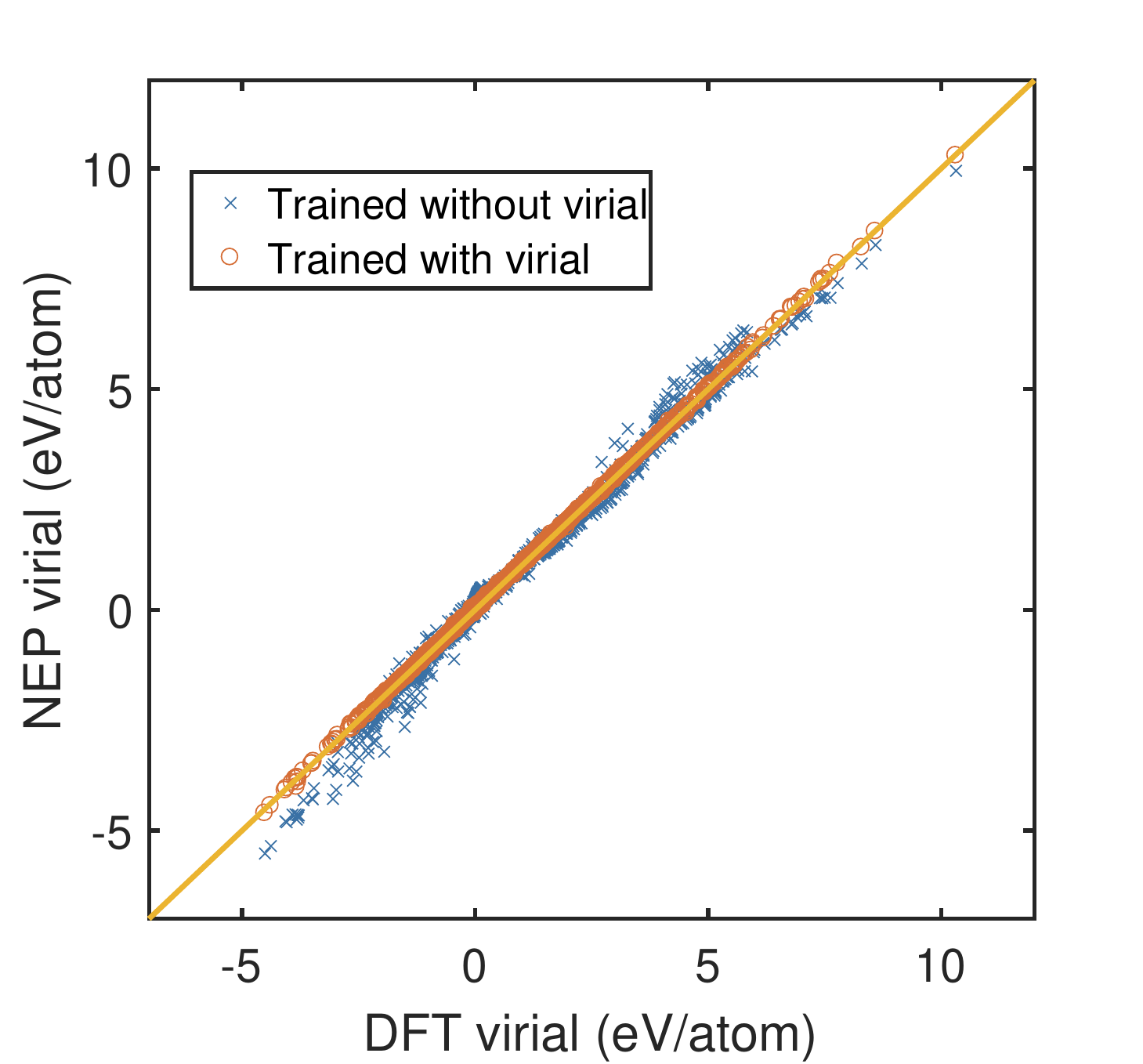}
\caption{
    Virial values from the NEP models in comparison with \gls{dft}-\gls{lda} data for the carbon data set. \cite{Deringer2017prb}
}
\label{fig:virial}
\end{figure}

Next we compute the zero-temperature elastic constants, for which the \verb"run.in" input file reads:
\begin{Verbatim}[frame=single]
potential  potentials/nep/C_2022_NEP3.txt 0
compute_elastic  0.01 cubic
\end{Verbatim}
Here, the \verb"compute_elastic" keyword is used to initiate the calculation of the three independent elastic constant components ($C_{11}$, $C_{12}$, and $C_{44}$) using the energy-strain relation with $\pm 1\%$ strain values.
The computed elastic constants are presented in \autoref{table:lattice_constant}.
The elastic constants from the NEP model are within 4\% of the reference \gls{dft}-\gls{lda} values.
For comparison, the $C_{12}$ value from \gls{gap} is about 24\% smaller than the reference \gls{dft}-\gls{lda} value.
We note that including virial information during training is crucial for obtaining accurate elastic properties, as can be seen from \autoref{fig:virial}.
In other words, fitting to energy and force data alone does not guarantee an accurate description of virials.
Since the calculation of the heat current involves virial terms, see Eq.~\eqref{equation:Ji}, this is also important for heat transport applications, as has already been pointed out by Shimamura \textit{et al.} \cite{shimamura2020jcp}.

\subsection{Tensile loading of diamond}

All the calculations above are static ones at zero temperature.
Here, we use \gls{md} simulations to study the fracture of diamond under uniaxial tensile loading.
The \verb"run.in" input file reads:
\begin{Verbatim}[frame=single]
potential potentials/nep/C_2022_NEP3.txt 0
velocity  300

ensemble  npt_ber 300 300 100 0 0 0 
          1000 1000 1000 1000
time_step 1
run       100000

ensemble      npt_scr 300 300 100 0 0 0 
              1000 1000 1000 1000
deform        1.42e-5 0 0 1
dump_thermo   100
dump_position 10000
run           5000000
\end{Verbatim}

\begin{figure}
\centering
\includegraphics[width=\columnwidth]{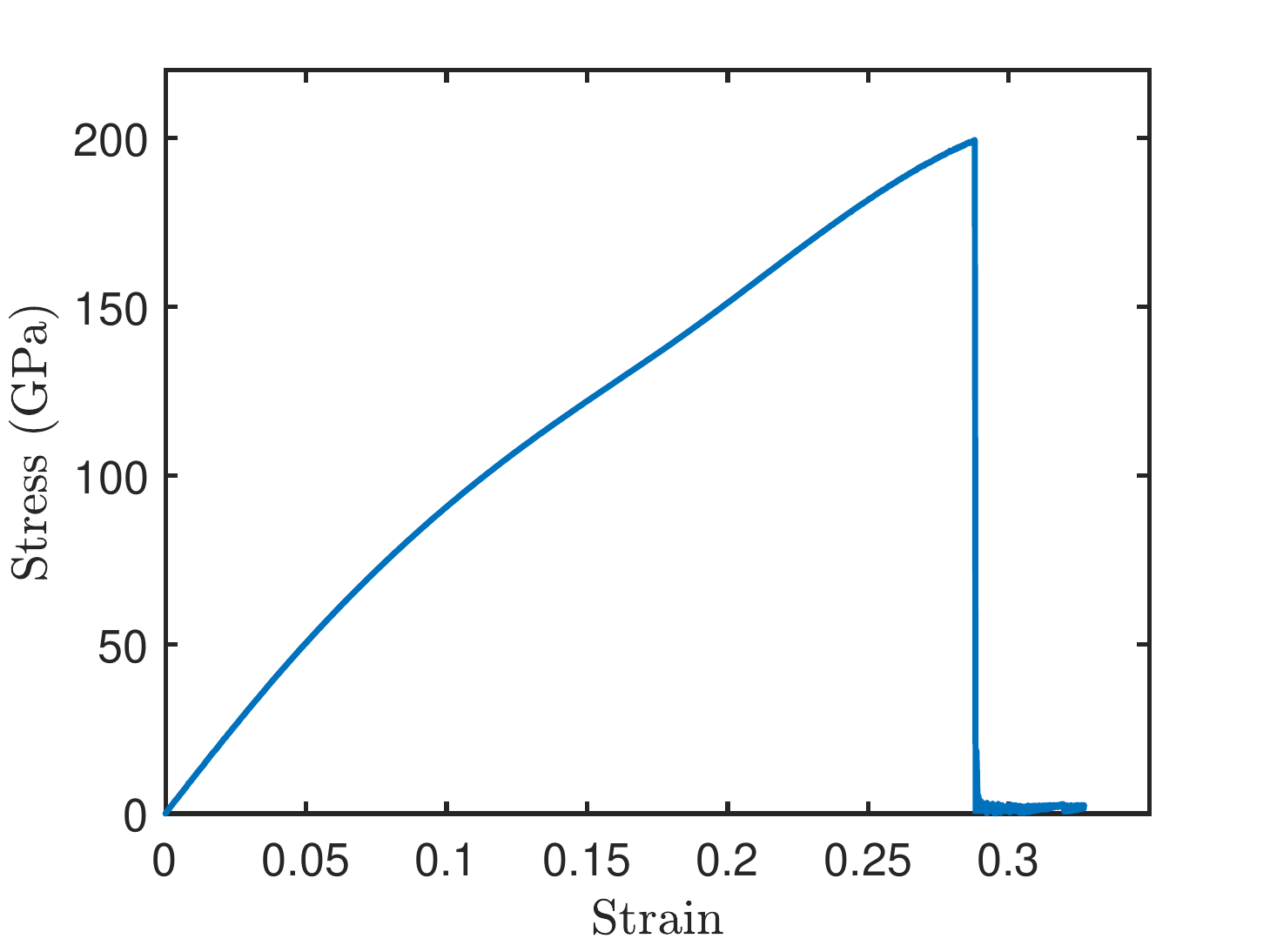}
\caption{
    Stress-strain relation from a uniaxial tensile loading simulation of diamond using the NEP (\SI{4.2}{\angstrom}) model for carbon.
}
\label{fig:stress_strain}
\end{figure}

\begin{figure}
\centering
\includegraphics[width=\columnwidth]{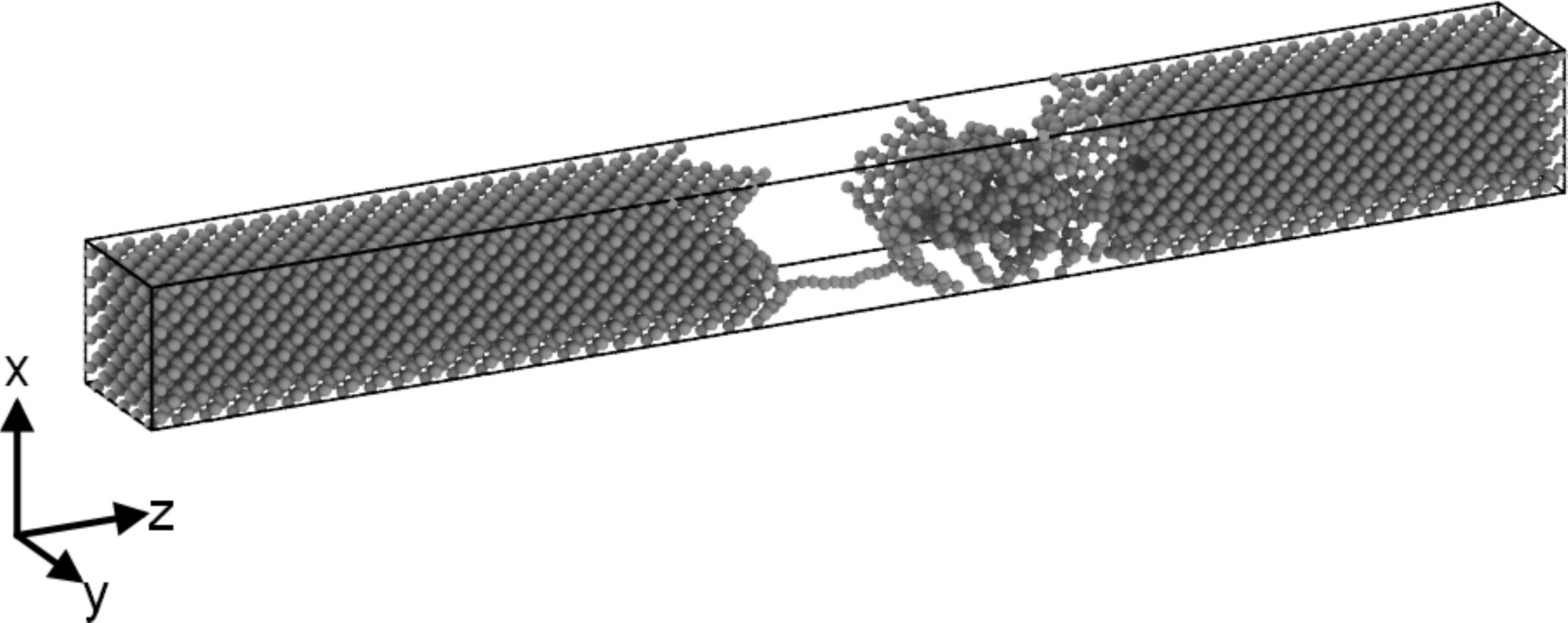}
\caption{
    A snapshot sampled during the fracture process under uniaxial tensile loading.
}
\label{fig:snapshot}
\end{figure}

The simulated model is a periodic supercell comprising of  $5\times 5 \times 40$ conventional cubic unit cells and hence 8,000 atoms.
We first equilibrate the system at \SI{300}{\kelvin} and zero pressure using the NPT ensemble via the Berendsen thermostat and barostat \cite{berendsen1984jcp} for \SI{100}{\pico\second}.
Then we switch to the Bussi-Donadio-Parrinello thermostat \cite{Bussi2007jcp} and the Bernetti-Bussi barostat \cite{Bernetti2020jcp} in the production stage, deforming the simulation box in the $z$ direction with a strain rate of \SI{1e8}{\per\second} for \SI{5}{\nano\second} up to a strain of 50\%, while the pressures in the $x$ and $y$ directions are set to zero.
Based on the output thermodynamic quantities and trajectory, we can obtain the stress-strain relation as shown in \autoref{fig:stress_strain} and identify a snapshot of the fracture process as shown in \autoref{fig:snapshot}.
The fracture is brittle with a fracture strength of about \SI{200}{\giga\pascal} at a strain of about 29\%.
We note that more independent simulations are needed to obtain statistically meaningful results beyond the current demonstration. 

\subsection{Quenching}

The carbon data set \cite{Deringer2017prb} is particularly suitable for studying liquid and amorphous carbon \cite{Deringer2017prb, caro2018prl, wang2022cm}.
In this example, we use a melt-quench-anneal protocol similar to that used in Ref.~\onlinecite{Deringer2017prb} (but with ten times longer simulation time for the relaxation at each temperature and an extra relaxation at 1000 K) to generate amorphous carbon.
The \verb"run.in" file reads:
\begin{Verbatim}[frame=single]
potential   potentials/nep/C_2022_NEP3.txt 0
velocity    9000

ensemble        nvt_lan 9000 9000 100
time_step       1
dump_thermo     10
dump_position   1000
run             30000  

ensemble        nvt_lan 5000 5000 100
dump_thermo     10
dump_position   1000
run             30000 

ensemble        nvt_lan 5000 1000 100
dump_thermo     10
dump_position   100
run             500

ensemble        nvt_lan 1000 1000 100
dump_thermo     10
dump_position   1000
run             30000 

ensemble        nvt_lan 300 300 100
dump_thermo     10
dump_position   1000
run             30000
\end{Verbatim}

\begin{figure*}
\centering
\includegraphics[width=2\columnwidth]{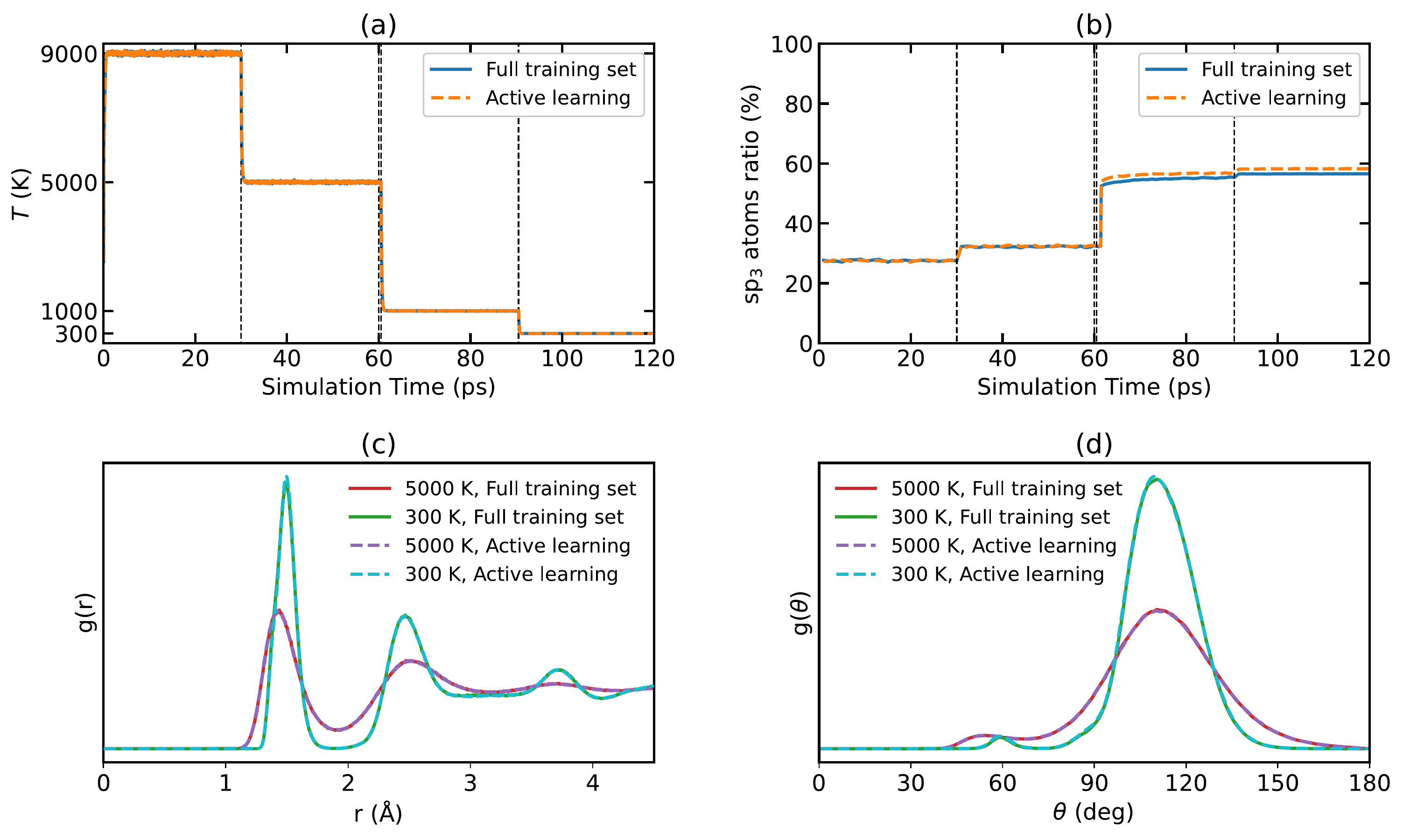}
\caption{
    (a) Temperature and (b) ratio of sp$^3$ bonded atoms as a function of simulation time as obtained using the NEP (\SI{4.2}{\angstrom}) model.
    (c) Radial and (d) angular distribution functions at \SI{5,000}{\kelvin} and \SI{300}{\kelvin}.}
\label{fig:quenching}
\end{figure*}

The initial simulation model is a cubic diamond supercell containing 64,000 atoms with a mass density of \SI{3.0}{\gram\per\centi\meter\cubed}.
The system is first quickly melted at \SI{9,000}{\kelvin} and then relaxed at \SI{5,000}{\kelvin}, followed by a quick quenching from \SI{5,000}{\kelvin} to \SI{1,000}{\kelvin} and further relaxation at \SI{1,000}{\kelvin} and \SI{300}{\kelvin}.
Here, the Langevin thermostat \cite{Bussi2007pre} is used to control the temperature.
The evolution of temperature and the ratio of sp$^3$-bonded atoms as a function of simulation time are presented in \autoref{fig:quenching}a-b.
The radial and angular distribution functions $g(r)$ and $g(\theta)$ at \SI{5,000}{\kelvin} and \SI{300}{\kelvin} in \autoref{fig:quenching}(c)-(d) show that the system is in liquid and amorphous-solid states, respectively.
We also performed the same \gls{md} simulation using the NEP model trained with the \gls{al} scheme in \autoref{section:active-learning} and we can see that it gives almost identical results as those from the NEP model trained using the full training data set.
This further demonstrates the effectiveness of our \gls{al} scheme based on the latent space.

\subsection{Density of states and heat capacity of amorphous carbon}

After obtaining amorphous carbon structures, we further study their thermal properties.
In this example, we calculate the \gls{vdos} for an amorphous carbon structure and then obtain the heat capacity with quantum corrections.
The \verb"run.in" file reads:
\begin{Verbatim}[frame=single]
potential   potentials/nep/C_2022_NEP3.txt 0
velocity    300

ensemble    nvt_ber 300 300 100
time_step   1
run         10000

ensemble    nve 
compute_dos 5 200 400
run         100000
\end{Verbatim}

The \gls{vdos} $\rho(\omega)$ is calculated from the \gls{vacf} \cite{dickey1969pr}.
The \gls{vacf} and \gls{vdos} are normalized to $3N$, where $N$ is the number of atoms.
The per-atom heat capacity with quantum corrections at the temperature $T$ is then calculated as
\begin{equation}
    C(T) = \frac{1}{N}\int_0^{\infty}
    \frac{d\omega}{2\pi} \rho(\omega) \frac{x^2 e^x}{(e^x-1)^2},
    \label{eq:heat-capacity-with-quantum-corrections}
\end{equation}
where $x=\hbar \omega / k_\mathrm{B} T$ is the ratio of the vibrational energy $\hbar \omega$ and the thermal energy $k_\mathrm{B} T$. 

\begin{figure}
\centering
\includegraphics[width=\columnwidth]{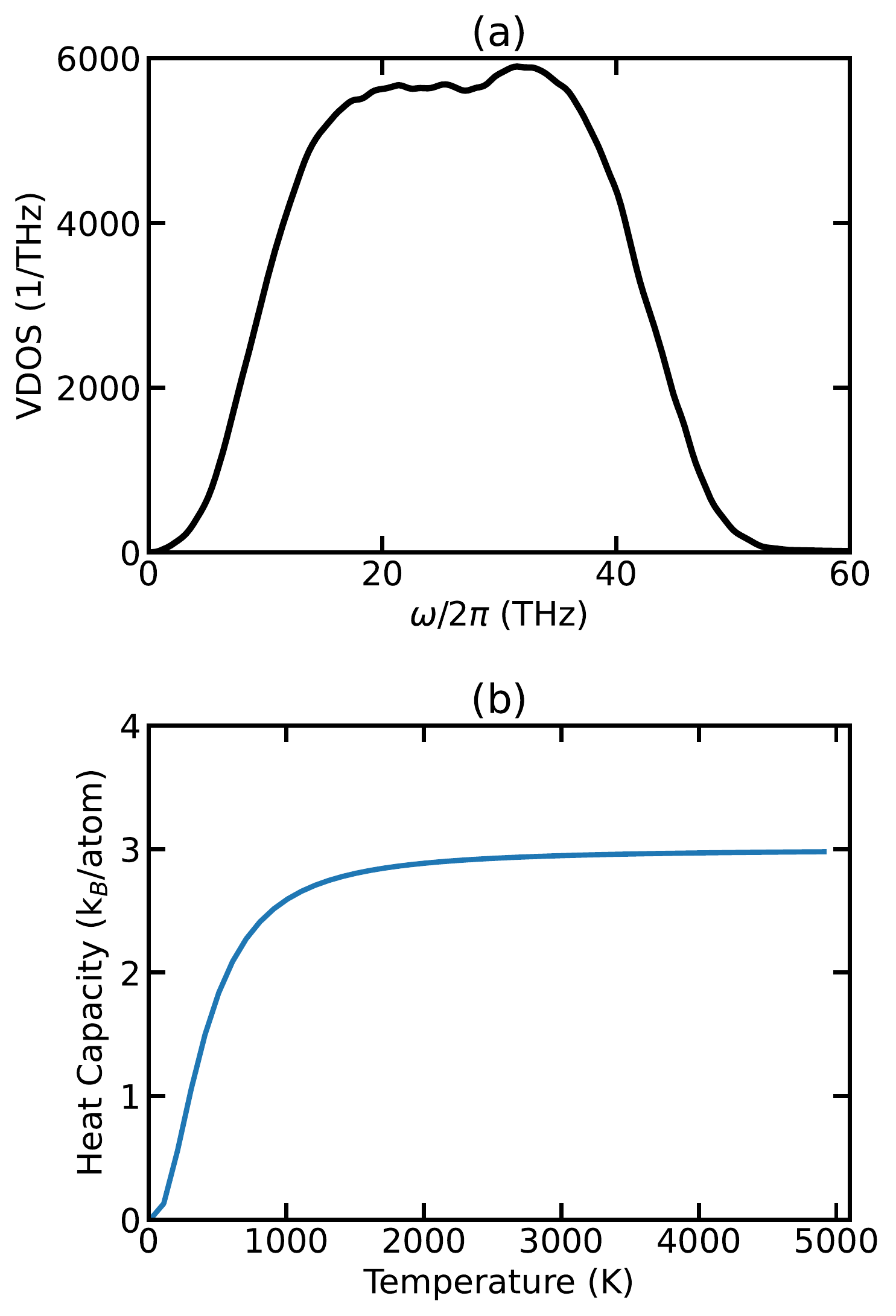}
\caption{
    (a) \Gls{vdos} of amorphous carbon with a density of \SI{3.0}{\gram\per\centi\meter\cubed} at \SI{300}{\kelvin} as obtained using the NEP (\SI{4.2}{\angstrom}) model.
    (b) Heat capacity with quantum corrections calculated from the \gls{vdos} via Eq.~\eqref{eq:heat-capacity-with-quantum-corrections}.
}
\label{fig:dos}
\end{figure}

\subsection{Thermal conductivity of amorphous carbon}

In this last example, we calculate the thermal conductivity of our amorphous carbon sample using the \gls{hnemd} method and the related spectral decomposition method \cite{fan2019prb, Gabourie2021}.
The \verb"run.in" file reads:
\begin{Verbatim}[frame=single]
potential   potentials/nep/C_2022_NEP3.txt 0
velocity    300

ensemble        nvt_nhc 300 300 100
time_step       1
run             100000

ensemble        nvt_nhc 300 300 100
compute_hnemd   1000 0 0 2e-4
compute_shc     5 200 2 500 400
run             2000000
\end{Verbatim}

\begin{figure}
\centering
\includegraphics[width=\columnwidth]{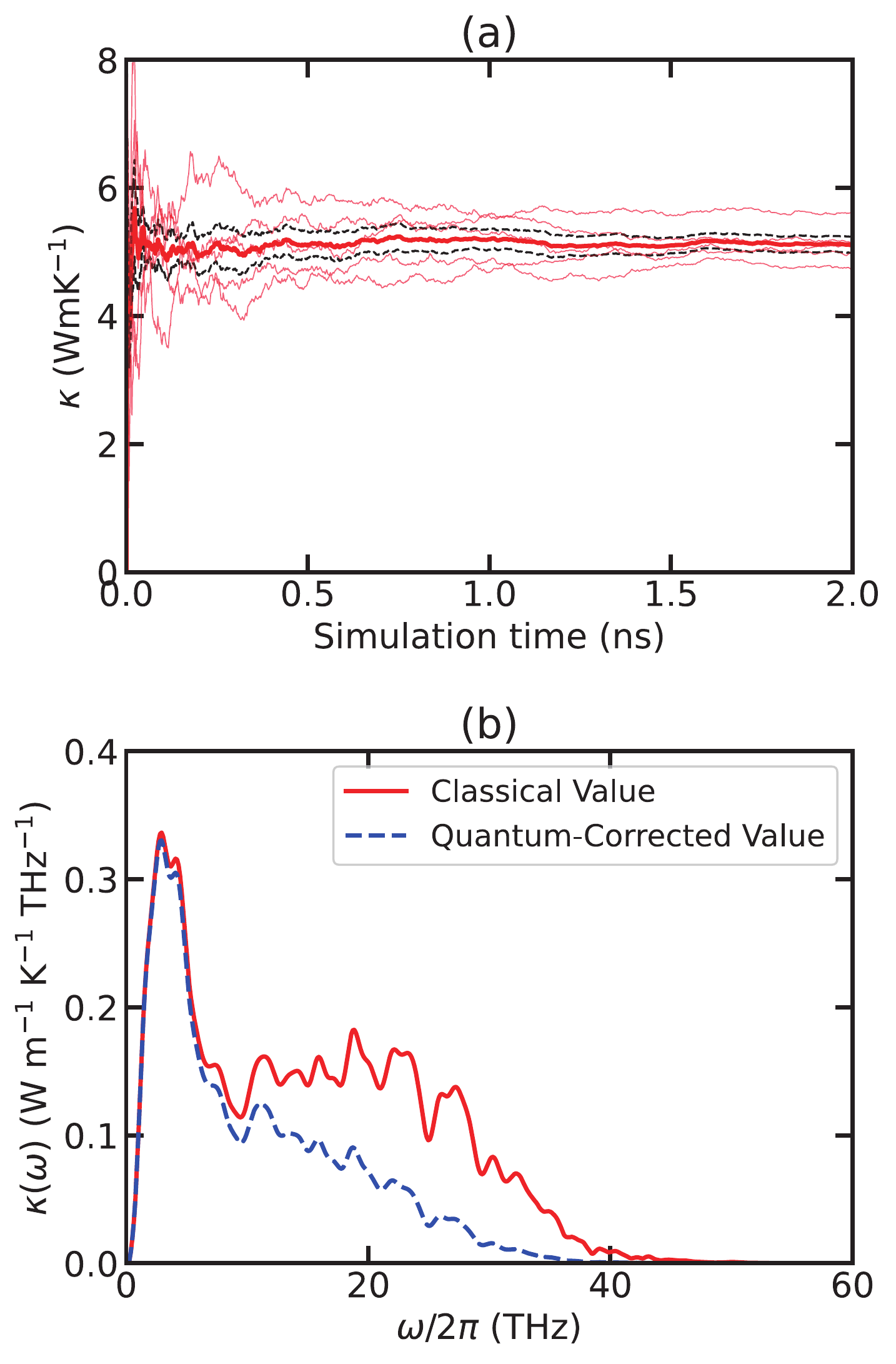}
\caption{
    (a) Thermal conductivity of amorphous carbon as calculated from the \gls{hnemd} method using the NEP (\SI{4.2}{\angstrom}) model.
    The thin solid lines are from 5 independent runs and the thick solid line is their average.
    The dashed lines represent the error bounds.
    (b) Classical and quantum-corrected spectral thermal conductivity as a function of the vibrational frequency.
}
\label{fig:hnemd}
\end{figure}

Thermal conductivity calculations usually require a lot of data to reduce the statistical uncertainty.
To this end, we perform a number of independent runs using the above inputs.
In \textsc{gpumd}, the velocities are automatically initialized with different pseudo-random number seeds for different runs.
Using the efficient \gls{hnemd} method, 5 independent runs (each with a production time of \SI{2}{\nano\second}) are sufficient to achieve high accuracy (small error bounds), as can be seen from \autoref{fig:hnemd}a.
The thermal conductivity of the amorphous carbon structure (with a mass density of \SI{3.0}{\gram\per\centi\meter\cubed}) at \SI{300}{\kelvin} is determined to be \SI{5.1 +- 0.1}{\watt\per\meter\per\kelvin}, where the statistical error is calculated as the standard error \cite{haile1992book}.
The thermal conductivity calculated in this way is the classical value.
For disordered materials, the thermal conductivity can be quantum corrected in a way similar to the quantum correction of the heat capacity \cite{gu2021jap}.
To achieve this, we first calculate the classical spectral thermal conductivity \cite{fan2019prb, Gabourie2021} $\kappa^\mathrm{c}(\omega)$ and include quantum corrections to obtain $\kappa^\mathrm{q}(\omega)$ as follows:
\begin{equation}
    \kappa^\mathrm{q}(\omega) = \kappa^\mathrm{c}(\omega) \frac{x^2 e^x}{(e^x-1)^2},
\end{equation}
where $x=\hbar \omega / k_\mathrm{B} T$.
Both $\kappa^\mathrm{c}(\omega)$ and $\kappa^\mathrm{q}(\omega)$ at \SI{300}{\kelvin} are shown in \autoref{fig:hnemd}b.
The quantum-corrected thermal conductivity at \SI{300}{\kelvin} is \SI{3.2 +- 0.1}{\watt\per\meter\per\kelvin}.
A more systematic investigation of the thermal transport properties in disordered carbon systems will be presented elsewhere.

\section{Interface to other codes}

A few Python packages have been developed to work with \textsc{gpumd} and are briefly presented below. See the \textit{Code availability} statement for the links of codes and documentations.

\subsection{The \textsc{gpyumd} package}

To help \textsc{gpumd} users generate input and process output files, we have developed a Python interface implemented in the \textsc{gpyumd} package.
Reading, preparing, and writing \verb"xyz.in" files is facilitated by the \verb"GpumdAtoms" class.
This class extends the \verb"Atoms" class from the popular Atomic Simulation Environment (\textsc{ase}) Python package \cite{Hjorth_Larsen_2017} to include \textsc{gpumd}-specific properties.
A simple example of writing an \verb"xyz.in" file is as follows:
\begin{lstlisting}[language=Python]
from ase.lattice.cubic import Diamond
from gpyumd.atoms import GpumdAtoms

Si = GpumdAtoms(Diamond("Si",size=(10,10,10)))
Si.set_max_neighbors(4)
Si.set_cutoff(3)
Si.write_gpumd() 
\end{lstlisting}

The \verb"GpumdAtoms" class also supports adding grouping methods, sorting atoms by group or type, generating \verb"basis.in" and \verb"kpoints.in" files for phonon calculations, and more.

The \textsc{gpyumd} package also has a \verb"Simulation" class that can be used to generate valid \verb"run.in" files.
To do so, in addition to checking each keyword parameter, it verifies that group selections in each keyword, atom types in potential definitions, and atom ordering in the \verb"xyz.in" file are consistent.
As a simple demonstration, continuing from our previous code snippet, we can create a simple \verb"run.in" file as follows:
\begin{lstlisting}[language=Python]
import gpyumd.keyword as kwd
from gpyumd.sim import Simulation

sim = Simulation(Si)
run = sim.add_run()
keywords = [
    kwd.Velocity(initial_temperature=300),
    kwd.TimeStep(dt_in_fs=1),
    kwd.Ensemble(ensemble_method='nve'),
    kwd.DumpThermo(interval=1000),
    kwd.RunKeyword(number_of_steps=1e5)]

[run.add_keyword(x) for x in keywords]
potential = kwd.Potential(filename='Si.txt', symbols=['Si'])
sim.add_potential(potential)
sim.create_simulation()
\end{lstlisting}

The output files of \textsc{gpumd} can be read and processed using simple \textsc{gpyumd} functions such as \verb"load_thermo()" for the \verb"thermo.out" file, \verb"load_hac()" for the \verb"hac.out" file, etc.
These functions return the data in convenient formats for data exploration in interactive environments such as Jupyter Notebooks.

\subsection{The \textsc{calorine} package}

To provide a deeper integration of \textsc{gpumd} within a Python workflow, we also provide the Python package \textsc{calorine}.
This section provides some examples for the functionality of this package.
A full documentation including extended examples and tutorials can be found at \url{https://calorine.materialsmodeling.org/}.

\subsubsection{ASE calculator}
The \textsc{calorine} package extends the functionality of \textsc{ase}, implementing an \textsc{ase} \texttt{Calculator} class, which lets users calculate energies, forces, and stresses with a NEP model directly from Python.
Under the hood, this calculator writes and reads the \textsc{gpumd} input and output files.
A minimal script reads as follows:
\begin{lstlisting}[language=Python]
from ase.build import bulk
from calorine import GPUNEP

calculator = GPUNEP('nep.txt')
atoms = bulk('Au', a=4.1)
atoms.set_calculator(calculator)
e = atoms.get_potential_energy()
\end{lstlisting}

This approach can, for example, greatly simplify the calculation of a large number of pre-defined structures, and provides access to various complex structural relaxation schemes available in \textsc{ase}.

\subsubsection{Interface to GPUMD}

\textsc{Calorine} also interfaces directly to \textsc{gpumd} using a \textsc{PyBind11} C++ interface.
This enables easy access to the data structures associated with the NEP implementation in \textsc{gpumd}.
At the moment the interface exposes a function for calculating the descriptors for an \textsc{ase} \texttt{Atoms} object, as well as a CPU-only \textsc{ase} \texttt{Calculator}. 
The CPU-only calculator enables using a trained NEP model on computer systems without a GPU. 
An example script for accessing the \gls{nep} descriptors for a structure and using the CPU-only calculator is given below.
\begin{lstlisting}[language=Python]
from ase import Atoms
from calorine.nepy import \
    get_descriptors, CPUNEP

atoms = Atoms('CO',
      positions=[[0, 0, 0], [0, 0, 1.1]],
      cell=[20, 20, 20])
descriptors = get_descriptors(atoms)

calculator = CPUNEP('nep.txt')
atoms.set_calculator(calculator)
e = atoms.get_potential_energy()
\end{lstlisting}

\subsection{The \textsc{pynep} package}

We also developed a Python package \textsc{pynep} to facilitate the \gls{al} process with a pre-trained \gls{nep} model.
It as well provides an \textsc{ase} \texttt{Calculator} to calculate the properties of an \texttt{Atoms} object, including energy, forces, stress, descriptors, and latent descriptors. A simple example script for calculating these properties is as follows:

\begin{lstlisting}[language=Python]
from ase.build import bulk
from pynep.calculate import NEP

# get energy and forces
calc = NEP('nep.txt')
atoms = bulk('C', 'diamond', cubic=True)
atoms.set_calculator(calc)
energy = atoms.get_potential_energy()
forces = atoms.get_forces()
stress = atoms.get_stress() 

# get descriptors and latent descriptors
des = calc.get_property('descriptor', atoms)
lat = calc.get_property('latent', atoms)
\end{lstlisting}

With the descriptors or the latent descriptors available, we can select structures with different sampling methods. An example script for selecting structures using the farthest-point sampling is given below: 

\begin{lstlisting}[language=Python]
from pynep.select import FarthestPointSample
from pynep.io import load_nep, dump_nep
import numpy as np

raw = load_nep('raw.in')
lat = np.array([np.mean(calc.get_property('latent', atoms), axis=0) for atoms in raw])
sampler = FarthestPointSample(min_distance=0.05)
selected = [raw[i] for i in sampler.select(lat, [])]
dump_nep('selected.in', selected)
\end{lstlisting}

\section{Summary and conclusions}

In summary, we have presented and reviewed the various features of the open-source \textsc{gpumd} package, with a focus on recent developments that have enabled the generation and use of accurate and efficient NEP \glspl{mlp} \cite{fan2021neuroevolution, fan2022jpcm}.
Two improvements on the atomic-environment descriptor have been introduced: one is to change the radial functions from Chebyshev basis functions to linear combinations of the basis functions, and the other is to extend the angular descriptor components by considering some 4-body and 5-body contributions as in the \gls{ace} approach \cite{drautz2019prb}.
Both of these extensions are shown to improve the accuracy of NEP models further.

We have used a diverse set of materials to demonstrate that the NEP approach can achieve an above-average accuracy compared to many other state-of-the-art \glspl{mlp}.
In addition, NEP models can achieve a far superior computational efficiency: typical NEP models are more than one order of magnitude faster than other \glspl{mlp} and similarly more memory efficient.
The high efficiency of NEP models originates from many aspects, including a reasonably small descriptor dimension (usually smaller than 100) combined with a simple neural network model with a single hidden layer, carefully derived expressions of the descriptor components, a balanced choice of the radial and angular cutoff distances, and finally a carefully optimized GPU implementation.
We present our algorithms in detail in Appendix~\ref{section:algorithms}.
The latent space in the simple neural network model of NEP models also allows us to construct an effective \gls{al} scheme that can be used to greatly reduce the computational efforts in the preparation of training data. 

Apart from being highly efficient, \textsc{gpumd} is also user-friendly.
It can both be used as a standalone package and be integrated with other packages such as \textsc{ase} \cite{Hjorth_Larsen_2017} via the \textsc{gpyumd}, \textsc{calorine}, and \textsc{pynep} Python packages. The use of an efficient derivative-free optimization algorithm (\gls{snes}) greatly simplifies the implementation and excludes the dependence of \textsc{gpumd} on any third-party machine-learning libraries, making the installation of \textsc{gpumd} effortless.

Finally, the NEP models trained using the \verb"nep" executable can be directly used by the \verb"gpumd" executable to perform atomistic simulations of various materials properties.
To demonstrate the range of properties, length and time scales that can be accessed via this approach, we have presented a series of examples using a NEP model trained using a standard carbon data set \cite{Deringer2017prb}. 

One of the disadvantages of \textsc{gpumd} is that it is still not very feature-rich (as compared to similar packages such as \textsc{lammps} \cite{thompson2022cpc}).
However, its open-source nature and the well-designed GPU-acceleration framework have been attracting more and more developers with diverse backgrounds who are enriching the features of \textsc{gpumd} at a fast pace.

\vspace{0.5cm}
\noindent{\textbf{Data availability}}

The training and testing results using the various \glspl{mlp} as presented in \autoref{section:performance} and \autoref{section:active-learning} are freely available via Zenodo \cite{zenodo_data}.
The input and output files for the \textsc{gpumd} examples presented in \autoref{section:examples} are included in the \textsc{gpumd} package (\url{https://github.com/brucefan1983/GPUMD}).

\vspace{0.5cm}
\noindent{\textbf{Code availability}}

The source code and documentation for \textsc{gpumd} are available at \url{https://github.com/brucefan1983/GPUMD} and \url{https://gpumd.zheyongfan.org/}, respectively.

The source code and documentation for \textsc{gpyumd} are available at \url{https://github.com/AlexGabourie/gpyumd} and \url{https://gpyumd.readthedocs.io/}, respectively. 

The source code and documentation for \textsc{calorine} are available at  \url{https://gitlab.com/materials-modeling/calorine} and \url{https://calorine.materialsmodeling.org/}, respectively.

The source code and documentation for \textsc{pynep} are available at \url{https://github.com/bigd4/PyNEP} and \url{https://pynep.readthedocs.io/}, respectively.

\vspace{0.5cm}
\noindent{\textbf{Conflict of Interest}}

The authors have no conflicts to disclose. 

\begin{acknowledgments}
Z.F. acknowledges support from the National Natural Science Foundation of China (NSFC) (No. 11974059).
Y.W., K.S., J.L., and H.D. acknowledge the support from the National Key Research and Development Program of China (2018YFB0704300).
The work at Nanjing University (J.J.W, Y.W., J.S.) is partially supported from the NSFC (grant nos. 12125404, 11974162, and 11834006), and the Fundamental Research Funds for the Central Universities.
The calculations performed in Nanjing University were carried out using supercomputers at the High Performance Computing Center of Collaborative Innovation Center of Advanced Microstructures, the high-performance supercomputing center of Nanjing University.
P.Y. and Z.Z. acknowledge support from the NSFC (No. 11932005).
T.A-N. has been supported in part by the Academy of Finland through its QTF Centre of Excellence program (No. 312298) and Technology Industries of Finland Centennial Foundation Future Makers grant.
Z.Z. and Y.C. are grateful for the research computing facilities offered by ITS, HKU.
J.M.R., E.L., and P.E. acknowledge support from the Swedish Research Council (2018-06482, 2020-04935, 2021-05072) and the Swedish Foundation for Strategic Research (SSF) via the SwedNess program (GSn15-0008) as well as computational resources provided by the Swedish National Infrastructure for Computing (SNIC) at NSC, C3SE, and PDC partially funded by the Swedish Research Council (2018-05973).
\end{acknowledgments}

\appendix

\section{Some derivations on the heat current expressions}
\label{section:heat_current}

Here we show that the heat current expressions for the multi-body potentials as considered in Ref.~\onlinecite{Boone2019jctc} are special cases of the general expressions in our formulation.
Without loss of generality, we take the 3-body potential as an example to show this.

Using the chain rule, we first rewrite Eq.~\eqref{equation:J-2} as
\begin{equation}
    \label{equation:J-3}
    \bm{J} = - \sum_i \sum_{j \neq i} \bm{r}_{ij} \frac{\partial U_i}{\partial \bm{r}_{j}} \cdot \bm{v}_j.
\end{equation}
Boone \textit{et al.} \cite{Boone2019jctc} considered explicit $m$-body potentials ($m=2,3,4$) that are usually used in topological force fields for organics.
For the 3-body potential considered therein, the site potential $U_i$ of atom $i$ is taken as the average of the potentials of the triplets it belongs to (the first index denotes the central atom of a triplet):
\begin{equation}
\label{equation:Ui-3body}
    U_i = \frac{1}{3} \sum_{k\neq i}\sum_{l\neq i} \left(U_{ikl}+U_{kli}+U_{lik}\right).
\end{equation}
Substituting Eq.~\eqref{equation:Ui-3body} into Eq.~\eqref{equation:J-3}, we have
\begin{equation}
    \label{equation:J3}
    \bm{J} = - \frac{1}{3}\sum_i \sum_{j \neq i} \sum_{k\neq i}\sum_{l\neq i} \bm{r}_{ij} \frac{\partial \left(U_{ikl}+U_{kli}+U_{lik}\right)}{\partial \bm{r}_{j}} \cdot \bm{v}_j.
\end{equation}
Note that $j$ cannot be $i$ but could be $k$ or $l$ and we thus have
\begin{equation}
    \frac{\partial U_{ikl}}{\partial \bm{r}_{j}} = \delta_{jk} \frac{\partial U_{ijl}}{\partial \bm{r}_{j}} + \delta_{jl} \frac{\partial U_{ikj}}{\partial \bm{r}_{j}}
\end{equation}
and similar expressions for ${\partial U_{kli}}/{\partial \bm{r}_{j}}$ and ${\partial U_{lik}}/{\partial \bm{r}_{j}}$.
Following Boone \textit{et al.} \cite{Boone2019jctc} we can define
\begin{equation}
\bm{F}_{j}^{ijl} \equiv - \frac{\partial U_{ijl}}{\partial \bm{r}_{j}}.
\end{equation}
as the force acting on atom $j$ from the triplet $ijl$. Then we can write Eq. (\ref{equation:J3}) as 
\begin{align}
\bm{J} 
&= \frac{1}{3}\sum_i \sum_{j \neq i} \sum_{k\neq i} \bm{r}_{ij} \nonumber \\
&\left(\bm{F}_{j}^{ijk} + \bm{F}_{j}^{ikj} + \bm{F}_{j}^{jki} + \bm{F}_{j}^{kji} + \bm{F}_{j}^{jik} + \bm{F}_{j}^{kij} \right) \cdot \bm{v}_j.    
\end{align}
By manipulating the dummy indices in the summation, it can be written as
\begin{align}
    \bm{J}
    &= \frac{1}{3}\sum_i \sum_{j \neq i} \sum_{k\neq i} 
    \left(\bm{r}_{ji} + \bm{r}_{ki}\right) \bm{F}_{i}^{ijk} \cdot \bm{v}_i \nonumber \\
    &+ 
    \left(\bm{r}_{ij} + \bm{r}_{kj}\right) \bm{F}_{j}^{ijk} \cdot \bm{v}_j +
    \left(\bm{r}_{ik} + \bm{r}_{jk}\right) \bm{F}_{k}^{ijk} \cdot \bm{v}_k,
\end{align}
which corresponds to Eq.~(18d) in Ref.~\onlinecite{Boone2019jctc}.

\section{Algorithms\label{section:algorithms}}

In this Appendix, we present the complete algorithms for evaluating the NEP energy, force, and virial expressions as implemented in \textsc{gpumd}.
First we list all the relevant quantities:
\begin{enumerate}
\item $N$ is the total number of atoms.
\item $\mathrm{NN}_i^\mathrm{R}$ is the number of neighbors of atom $i$ for the radial descriptor components.
\item $\mathrm{NN}_i^\mathrm{A}$ is the number of neighbors of atom $i$ for the angular descriptor components.
\item $\mathrm{NL}_{im}^\mathrm{R}$ is the index of the $m^{\rm th}$ neighbor of atom $i$ for the radial descriptor components.
\item $\mathrm{NL}_{im}^\mathrm{A}$ is the index of the $m^{\rm th}$ neighbor of atom $i$ for the angular descriptor components.
\item $\{\bm{r}_i\}_{i=0}^{N-1}$ are the atom positions.
\item $\{U_i\}_{i=0}^{N-1}$ are the site energies.
\item $\{\partial U_i/\partial q^i_{\nu}\}_{i=0}^{N-1}$ are the derivatives of the site energies with respect to the descriptor components.
\item $\{\partial U_i/\partial \bm{r}_{ij}\}_{i=0}^{N-1}$ are the partial forces.
\item $\{\bm{F}_i\}_{i=0}^{N-1}$ are the forces on the atoms.
\item $\{\mathbf{W}_i\}_{i=0}^{N-1}$ are the virials on the atoms.
\end{enumerate}

\subsection{Application of the neural network}
\label{section:ann}

We use a simple feedforward neural network with a single hidden layer.
We have done extensive tests and found that a single hidden layer is sufficient to achieve high accuracy for NEP models and using more hidden layers only reduces computational performance without a corresponding improvement in model accuracy.
Using a single hidden layer, many variables can be defined as registers instead of global memory or local memory in the CUDA kernel, which are much more expensive to access.
Therefore, using a single hidden layer can achieve a significantly higher computational performance in the parallelism scheme we adopted. 

\begin{lstlisting}[language=C++,caption={The function applying the feedforward neural network to the input descriptor vector to obtain the site energy of an atom and the derivative of the energy with respect to the descriptor components. },label={listing:apply_ann}]
__device__ void apply_ann(
  const int N_des,
  const int N_neu,
  const float* w0,
  const float* b0,
  const float* w1,
  const float* b1,
  const float* q,
  float& energy,
  float* energy_derivative)
{
  for (int n = 0; n < N_neu; ++n) {
    float w0_times_q = 0.0f;
    for (int d = 0; d < N_des; ++d) {
      w0_times_q += w0[n*N_des+d] * q[d];
    }
    float x1 = tanh(w0_times_q - b0[n]);
    float tanh_der = 1.0f - x1 * x1;
    energy += w1[n] * x1;
    for (int d = 0; d < N_des; ++d) {
      float y1 = tanh_der * w0[n*N_des+d];
      energy_derivative[d] += w1[n] * y1;
    }
  }
  energy -= b1[0];
}
\end{lstlisting}

The complete \verb"__device__" function applying the neural network is presented in \autoref{listing:apply_ann}.
For the inputs, \verb"N_des" is the dimension $N_\mathrm{des}$ of the descriptor vector, \verb"N_neu" is the number of neurons $N_\mathrm{neu}$ in the hidden layer, \verb"w0" is the weight matrix $\mathbf{w}^{(0)}$, \verb"w1" is the weight vector $\mathbf{w}^{(1)}$, \verb"b0" is the bias vector $\mathbf{b}^{(0)}$ in the hidden layer, \verb"b1" is the bias $b^{(1)}$ in the output node, and \verb"q" is the descriptor vector $\mathbf{q}^i$.
For the outputs, \verb"energy" is the site energy $U_i$ and \verb"energy_derivative" is the derivative of the site energy with respect to the descriptor components $\partial U_i/\partial q_{\nu}^i$.
Note that we do not need to calculate the derivative of the energy (and other related quantities such as force and virial) with respect to the neural-network parameters, as required in the conventional gradient-descent approach.
In our evolutionary algorithm approach, there is no need to calculate the derivative of the loss function with respect to any parameters.
Therefore, our implementation is very simple regarding the neural network part and particularly, we do not make \textsc{gpumd} dependent on any third-party packages.
This makes the installation of \textsc{gpumd} very simple and straightforward. 

\subsection{Energy and derivative of energy with respect to descriptor}

In the first CUDA kernel (see \autoref{algo:energy}), the thread associated with atom $i$ calculates the whole descriptor vector $\mathbf{q}^i$ and calls the \verb"__device__" function in \autoref{listing:apply_ann} to obtain the energy $U_i$ and the derivatives $\partial U_i/\partial q^i_{\nu}$.
The derivatives will be used in the next two CUDA kernels.

\begin{algorithm}[htb!]
\DontPrintSemicolon 
Assign atom $i$ to CUDA thread $i$\;
\If{$i < N$}{
  Read position $\bm{r}_i$ for atom $i$ from global memory\;
  \For{$m=0$ to $\mathrm{NN}_i^\mathrm{R} - 1$ }{
    $j \leftarrow \mathrm{NL}_{im}^\mathrm{R}$ \;
    Read in $\bm{r}_j$ from global memory and calculate $\bm{r}_{ij}$ \;
    $\bm{r}_{ij} \leftarrow $ minimum image of $\bm{r}_{ij}$ \;
    Calculate the radial functions $g_n(r_{ij})$ according to Eq.~\eqref{equation:g_n} \;
    Accumulate the radial descriptor components according to Eq.~\eqref{equation:qin} \;
  }
  \For{$n=0$ to $n_\mathrm{max}^\mathrm{A}$}{
    \For{$m=0$ to $\mathrm{NN}_i^\mathrm{A} - 1$ }{
      $j \leftarrow \mathrm{NL}_{im}^\mathrm{A}$ \;
      Read in $\bm{r}_j$ from global memory and calculate $\bm{r}_{ij}$ \;
      $\bm{r}_{ij} \leftarrow $ minimum image of $\bm{r}_{ij}$ \;
      Calculate the radial functions $g_n(r_{ij})$ according to Eq.~\eqref{equation:g_n} but with $r_\mathrm{c}^\mathrm{R}$ changed to $r_\mathrm{c}^\mathrm{A}$\;
      Accumulate the summations $S_{n,k}$ according to Eq.~\eqref{equation:S_nk} \;
    }
    Calculate the angular descriptor components for the current $n$ according to the equations in \autoref{section:explicit_expressions}.\;
    Save the summations $S_{n,k}$ for the current $n$ to global memory for later use.\;
  }
  Apply the neural network model to get the energy $U_i$ and energy derivatives $\partial U_i/\partial q^i_{\nu}$ from the descriptor $q^i_{\nu}$ and save them to global memory.\;
}
\caption{Pseudo-code of the CUDA kernel for evaluating the descriptor vector $q^i_{\nu}$, the per-atom energy $U_i$, and the derivatives of the energy with respect the descriptor components $\partial U_i/\partial q^i_{\nu}$.}
\label{algo:energy}
\end{algorithm}

\subsection{Force and virial from the radial descriptor components}

In the second CUDA kernel (see \autoref{algo:radial_force}), the thread associated with atom $i$ first calculates the partial forces $\partial U_i/\partial \bm{r}_{ij}$ and $\partial U_j/\partial \bm{r}_{ji}$ related to the radial descriptor components, and then accumulates the force $\bm{F}_{i}$ and virial $\mathbf{W}_i$ on atom $i$.
For the radial descriptor components, $\partial U_i/\partial \bm{r}_{ij}$ and $\partial U_j/\partial \bm{r}_{ji}$ only differ a little and it is thus a good choice to calculate both within the CUDA kernel.
This algorithm is very similar to that for \gls{eam} potentials \cite{daw1984prb} (which is an angular-independent many-body potential) as implemented in \textsc{gpumd}.

\begin{algorithm}[htb!]
\DontPrintSemicolon 
Assign atom $i$ to CUDA thread $i$.\;
\If{$i < N$}{
  Read position $\bm{r}_i$ for atom $i$ from global memory\;
  \For{$m=0$ to $\mathrm{NN}_i^\mathrm{R} - 1$ }{
    $j \leftarrow \mathrm{NL}_{im}^\mathrm{R}$. \;
    Read in $\bm{r}_j$ from global memory and calculate $\bm{r}_{ij}$. \;
    $\bm{r}_{ij} \leftarrow $ minimum image of $\bm{r}_{ij}$. \;
    Calculate the partial force $\partial U_i/\partial \bm{r}_{ij}$ related to the radial descriptor components, i.e., the first term on the right hand side of Eq.~\eqref{equation:partial_force}. \;
    Similarly calculate the partial force $\partial U_j/\partial \bm{r}_{ji}$.\;
    Accumulate the force $\bm{F}_i$ on atom $i$ according to Eq.~\eqref{equation:F_ij} and Eq.~\eqref{equation:F_i}. \;
    Accumulate the virial $\mathbf{W}_i$ on atom $i$ according to Eq.~\eqref{equation:virial}. \;
  }
}
\caption{Pseudo-code of the CUDA kernel for evaluating the force and virial from the radial descriptor components.}
\label{algo:radial_force}
\end{algorithm}

\subsection{Partial forces from the angular descriptor components}

\begin{algorithm}[htb!]
\DontPrintSemicolon 
Assign atom $i$ to CUDA thread $i$\;
\If{$i < N$}{
  Read position $\bm{r}_i$ for atom $i$ from global memory\;
  \For{$m=0$ to $\mathrm{NN}_i^\mathrm{A} - 1$ }{
      $j \leftarrow \mathrm{NL}_{im}^\mathrm{A}$ \;
      Read in $\bm{r}_j$ from global memory and calculate $\bm{r}_{ij}$ \;
      $\bm{r}_{ij} \leftarrow $ minimum image of $\bm{r}_{ij}$ \;
      Calculate the partial force $\partial U_i/\partial \bm{r}_{ij}$ related to the angular descriptor components, i.e., the last three terms on the right hand side of  Eq.~\eqref{equation:partial_force}.\;
      Save the partial force $\partial U_i/\partial \bm{r}_{ij}$ to global memory for later use.
  }
}
\caption{Pseudo-code of the CUDA kernel for evaluating the partial forces $\partial U_i/\partial \bm{r}_{ij}$ for all atoms $i$ and all neighbors $j$ of $i$ from the angular descriptor components.}
\label{algo:angular_partial_force}
\end{algorithm}

In the third CUDA kernel (see \autoref{algo:angular_partial_force}), the thread associated with atom $i$ calculates the partial forces $\partial U_i/\partial \bm{r}_{ij}$ related to the angular descriptor components and saves them to global memory, which is then used in the next CUDA kernel.
For the angular descriptor components, $\partial U_i/\partial \bm{r}_{ij}$ and $\partial U_j/\partial \bm{r}_{ji}$ differ a lot and it is thus more efficient to use a two-kernel approach: using one CUDA kernel (the current one) to calculate the partial forces $\{\partial U_i/\partial \bm{r}_{ij}\}$ for all atom pairs (within the angular cutoff) and save them to global memory, and then using another CUDA kernel (the next one) to consume them. 

\subsection{Force and virial from the angular partial forces}

\begin{algorithm}[htb!]
\DontPrintSemicolon 
Assign atom $i$ to CUDA thread $i$.\;
\If{$i < N$}{
  Read position $\bm{r}_i$ for atom $i$ from global memory\;
  \For{$m=0$ to $\mathrm{NN}_i^\mathrm{A} - 1$ }{
    $j \leftarrow \mathrm{NL}_{im}^\mathrm{A}$. \;
    Read in $\bm{r}_j$ from global memory and calculate $\bm{r}_{ij}$. \;
    $\bm{r}_{ij} \leftarrow $ minimum image of $\bm{r}_{ij}$. \;
    Read in the partial force $\partial U_i/\partial \bm{r}_{ij}$ related to the angular descriptor components from global memory. \;
    Similarly read in $U_j/\partial \bm{r}_{ji}$ from global memory with some index manipulations.\;
    Accumulate the force $\bm{F}_i$ on atom $i$ according to Eq.~(\ref{equation:F_ij}) and Eq.~\eqref{equation:F_i}. \;
    Accumulate the virial $\mathbf{W}_i$ on atom $i$ according to Eq.~\eqref{equation:virial}. \;
  }
}
\caption{Pseudo-code of the CUDA kernel for evaluating the force and virial from the angular descriptor components.}
\label{algo:angular_force}
\end{algorithm}

After obtaining the partial force $\{\partial U_i/\partial \bm{r}_{ij}\}$ related to the angular descriptor components, we use a CUDA kernel (see \autoref{algo:angular_force}) to accumulate the corresponding force and virial.
In this kernel, we load the partial force $\partial U_i/\partial \bm{r}_{ij}$ and $\partial U_j/\partial \bm{r}_{ji}$, which are related to each other by an exchange of atom indices $i$ and $j$.
Then we accumulate the force $\bm{F}_i$ on atom $i$ according to Eqs.~\eqref{equation:F_ij} and \eqref{equation:F_i}, and accumulate the virial $\mathbf{W}_i$ on atom $i$ according to Eq.~\eqref{equation:virial}.
This is a general CUDA kernel used in the \textsc{gpumd} package for all the angular-dependent many-body potentials, such as the Stillinger-Weber \cite{stillinger1985prb} and Tersoff potentials \cite{tersoff1989prb}. 

%

\end{document}